\documentclass[preprint,showpacs,eqsecnum,aps,amsmath,amssymb,floatfix]{revtex4}
\usepackage{graphics}
\usepackage{epsfig}

\begin{document}

\title{Wobbling motion in $^{165,167}$Lu within a semi-classical framework}

\author{A. A. Raduta$^{a), b)}$,  R. Poenaru $^{a)}$ and Al. H. Raduta $^{a)}$ }

\address{$^{a)}$ Department of Theoretical Physics, Institute of Physics and
  Nuclear Engineering, Bucharest, POBox MG6, Romania}

\address{$^{b)}$Academy of Romanian Scientists, 54 Splaiul Independentei, Bucharest 050094, Romania}

\begin{abstract}
The results obtained for $^{165,167}$Lu with a semi-classical formalism are presented. Properties like excitation energies for the super-deformed bands TSD1, TSD2, TSD3, in $^{165}$Lu, and TSD1 and TSD2 for $^{167}$Lu, inter- and intra-band B(E2) and B(M1), the mixing ratios, transition quadrupole moments are compared either with the corresponding experimental data or with those obtained for $^{163}$Lu. Also alignments, dynamic moments of inertia,
relative energy to a reference energy of a rigid  symmetric rotor with an effective moment of inertia and the angle between the angular momenta of the core and odd nucleon were quantitatively studied. One concludes that the semi-classical formalism provides a realistic description of all known wobbling features in $^{165, 167}$Lu.
\end{abstract} 
\pacs{21.10.Re, 21.60.Ev,27.70.+q}
\maketitle
\section{Introduction}
\renewcommand{\theequation}{1.\arabic{equation}}
\setcounter{equation}{0}
The wobbling motion is considered to be a signature of the triaxial nuclei. Essentially such a motion consists in a precession of the total angular momentum of a triaxial system combined with an oscillation of its projection on the quantization axis around a steady position. The first suggestion of a wobbling motion was made by Bohr and Mottelson within a triaxial rotor  model for high spin states, in which the total angular momentum almost aligns to the principal axis with the largest moment of inertia \cite{BMott}. This event was followed by a fully microscopic description due to Marshalek \cite{Marsh}. Since these two reference contributions showed up a large volume of experimental and theoretical results has been accumulated
\cite{Odeg,Jens,Ikuko,Scho,Amro,Gorg,Ham,Matsu,Ham1,Jens1,Hage,Tana3,Oi,Bring,Hart,Cast,Alme,MikIans}. Experimentally, the wobbling states excited in triaxial super-deformed (TSD) bands are known in several nuclei like $^{161,163,165,167}$Lu and $^{167}$Ta \cite{Bring,Hart}.

In a previous publication \cite{Rad017} we formulated a semi-classical formalism as to describe the main features of the wobbling motion, which was successfully applied to $^{163}$Lu.
Here, we present the results obtained within the quoted formalism for $^{165}$Lu. Besides the excitation energies, intra- and inter-band transitions, new properties like
the alignment, the dynamic moment of inertia, the departure from the rigid rotor with an effective moment of inertia, the angle between angular momenta of the core and odd nucleon, were investigated.

The above sketched project will be presented according to the following plan.
For the sake of consistency of our exposure we briefly describe the main ingredients of the proposed formalism in Section 2. An instructive comparison of the present formalism with a boson expansion method is given in Section 3. Numerical application  for $^{165}$Lu is described in Section 4, while the final conclusions are summarized in Section 5.

\section{Brief review of the semi-classical approach}
\renewcommand{\theequation}{2.\arabic{equation}}
\setcounter{equation}{0}

Here we study an odd-mass nuclear system  consisting of an even-even core described by a triaxial rotor Hamiltonian $H_{rot}$ and a single j-shell particle moving in a quadrupole deformed mean-field:
\begin{equation}
H_{sp}=\frac{V}{j(j+1)}\left[\cos\gamma(3j_3^2-{\bf j}^2)-\sqrt{3}\sin\gamma(j_1^2-j_2^2)\right].
\label{hassp}
\end{equation}
In terms of the total angular momentum ${\bf I}(={\bf R}+{\bf j}) $ and the angular momentum carried by the odd particle, ${\bf j}$, the rotor Hamiltonian is written as:
\begin{equation}
H_{rot}=\sum_{k=1,2,3}A_k(I_k-j_k)^2.
\end{equation}
where $A_k$ are half of the reciprocal moments of inertia associated to the principal axes of the inertia ellipsoid i.e., $A_k=1/(2{\cal I}_k)$.

The moments of inertia are given by the rigid-body model \cite{Ring}:
\begin{equation}
{\cal I}^{rig}_{k}=\frac{{\cal I}_0}{1+(\frac{5}{16\pi})^{1/2}\beta}\left[1-\left(\frac{5}{4\pi}\right)^{1/2}\beta\cos\left(\gamma+\frac{2}{3}\pi k\right)\right],\;k=1,2,3
\end{equation}
The total Hamiltonian, $\hat{H}$, is dequantized through  the time dependent variational principle:
\begin{equation}
\delta\int_{0}^{t}\langle \Psi_{IjM}|{\hat H}-i\frac{\partial}{\partial t'}|\Psi_{IjM}\rangle d t'=0,
\end{equation}
with the trial function chosen as:
\begin{equation}
|\Psi_{Ij;M}\rangle ={\bf N}e^{z\hat{I}_-}e^{s\hat{j}_-}|IMI\rangle |jj\rangle ,
\end{equation} 
with $\hat{I}_-$ and $\hat{j}_-$ denoting the lowering operators for the intrinsic angular momenta ${\bf I}$ and ${\bf j}$ respectively, while ${\bf N}$ is  the normalization factor.
$|IMI\rangle $ and $|jj\rangle$ are extremal states for the operators ${\hat I}^2, {\hat I}_3$ and ${\hat j}^2, {\hat j}_3$, respectively.
 
The variables $z$ and $s$ are complex functions of time and play the role of classical phase space coordinates describing the motion of the core and the odd particle, respectively:
\begin{equation}
z=\rho e^{i\varphi},\;\;s=fe^{i\psi}.
\end{equation}
The new variables $(\varphi, r)$ and $(\psi,t)$ with $r$ and $t$ defined as:
\begin{equation}
r=\frac{2I}{1+\rho^2},\;\;0\le r\le 2I;\;\;
t=\frac{2j}{1+f^2},\;\; 0\le t\le 2j,
\end{equation}
bring the classical equations of motion, provided by the variational principle, to the canonical form:
\begin{eqnarray}
\frac{\partial {\cal H}}{\partial r}&=&\stackrel{\bullet}{\varphi};\;\frac{\partial {\cal H}}{\partial \varphi}=-\stackrel{\bullet}{r} \nonumber\\ 
\frac{\partial {\cal H}}{\partial t}&=&\stackrel{\bullet}{\psi};\;\frac{\partial {\cal H}}{\partial \psi}=-\stackrel{\bullet}{t}. 
\label{eqmot}
\end{eqnarray}
where ${\cal H}$ denotes the average of $\hat{H}$ with the trial function $|\Psi_{IjM}\rangle$ and plays the role of the classical energy function.
Equations of motion (2.8) are explicitly given in Appendix A.
The classical energy has the expression $^{[0]}$\setcounter{footnote}{0}\footnotetext{In the expression of ${\cal H}$ from Ref. \cite{Rad017} a lamentable error appeared. Indeed, in the second last line of Eq. (\ref{classen}) the factor 2 multiplying the square root term as well as the free term $-2A_3Ij$ are missing. Corrections for these propagate to the other equations in the following way: a) In the expressions of $k,k',\omega_1,\omega_2, B'$ and $C'$ the terms $jA_1$ and $IA_1$ are doubled. Also, in $B'$ and $C'$ the factors $A_2$ and $A_3$ are to be multiplied with 2. In C the terms $2IjA_2A_3,\; IjA_2^2$ and $IjA_3^2$ have to be multiplied by 4. We repeated the numerical calculations with the corrected formulas and found out that results are quantitatively preserved, except for the fitting parameters, ${\cal I}_0$ and $V$.}: 
\begin{eqnarray}
{\cal H}&\equiv&\langle \Psi_{IjM}|H|\Psi_{IjM}\rangle\nonumber\\
        &=&\frac{I}{2}(A_1+A_2)+A_3I^2+\frac{2I-1}{2I}r(2I-r)\left(A_1\cos^2\varphi+A_2\sin^2\varphi -A_3\right)\nonumber\\
        &+&\frac{j}{2}(A_1+A_2)+A_3j^2+\frac{2j-1}{2j}t(2j-t)\left(A_1\cos^2\psi+A_2\sin^2\psi -A_3\right)\nonumber\\
        &-&2\sqrt{r(2I-r)t(2j-t)}\left(A_1\cos\varphi\cos\psi+A_2\sin\varphi\sin\psi\right)+A_3\left(r(2j-t)+t(2I-r)\right)-2A_3Ij\nonumber\\
        &+&V\frac{2j-1}{j+1}\left[\cos\gamma-\frac{t(2j-t)}{2j^2}\sqrt{3}\left(\sqrt{3}\cos\gamma +\sin\gamma\cos2\psi\right)\right].
\label{classen}
\end{eqnarray} 
and is minimal in the point
$(\varphi,r)=(0,I);(\psi,t)=(0,j)$, when $A_1<A_2<A_3$.  For rigid moments of inertia the mentioned restriction is valid for any $\gamma$ satisfying the inequalities: $0<\gamma<\pi/3$.
Linearizing the equations of motion around the minimum point of ${\cal H}$, one obtains a harmonic motion for the system, with the frequency given by the equation:
\begin{equation}
\Omega^4+B\Omega^2+C=0,
\label{ecOm}
\end{equation}
where the the coefficients B and C have the expressions:
\begin{eqnarray}
-B&=&\left[(2I-1)(A_3-A_1)+2jA_1\right]\left[(2I-1)(A_2-A_1)+2jA_1\right]+8A_2A_3Ij\nonumber\\
 &+&\left[(2j-1)(A_3-A_1)+2IA_1+V\frac{2j-1}{j(j+1)}\sqrt{3}(\sqrt{3}\cos\gamma+\sin\gamma)\right]\nonumber\\
 &\times&\left[(2j-1)(A_2-A_1)+2IA_1+V\frac{2j-1}{j(j+1)}2\sqrt{3}\sin\gamma\right],\\
C&=&\left\{\left[(2I-1)(A_3-A_1)+2jA_1\right]\left[(2j-1)(A_3-A_1)+2IA_1+V\frac{2j-1}{j(j+1)}\sqrt{3}(\sqrt{3}\cos\gamma+\sin\gamma)\right]\right. \nonumber\\
&-&\left. 4IjA_3^2\right \}\nonumber\\
 &\times&\left\{\left[(2I-1)(A_2-A_1)+2jA_1\right]\left[(2j-1)(A_2-A_1)+2IA_1+V\frac{2j-1}{j(j+1)}2\sqrt{3}\sin\gamma\right]-4IjA_2^2\right\}.\nonumber\\
\label{BandC}
\end{eqnarray}
Alternatively, we may expand the classical energy function up to second order in the coordinates deviations from the mentioned minimum ($(\varphi',r');(\psi',t')$), and then quantizing the result by the association:
\begin{eqnarray}
&&\varphi'\to\hat{q};\;\;r'\to \hat{p};\;\;\;\;[\hat{q},\hat{p}]=i,\nonumber\\
&&\psi'\to\hat{q_1};\;\;t'\to \hat{p_1};\;\;[\hat{q_1},\hat{p_1}]=i,
\end{eqnarray}
one obtains a Hamiltonian, given analytically by Eq. (A.2), which describes
two coupled harmonic vibrations of energies:
\begin{eqnarray}
\omega_1&=&\left[(2I-1)(A_3-A_1)+2jA_1\right]^{1/2}\left[(2I-1)(A_2-A_1)+2jA_1\right]^{1/2},\nonumber\\
\omega_2&=&\left[(2j-1)(A_3-A_1)+2IA_1+V\frac{2j-1}{j(j+1)}\sqrt{3}\left(\sqrt{3}\cos\gamma+\sin\gamma\right)\right]^{1/2}\nonumber\\
        &\times&\left[(2j-1)(A_2-A_1)+2IA_1+V\frac{2j-1}{j(j+1)}2\sqrt{3}\sin\gamma\right]^{1/2}.
\label{o1o2}
\end{eqnarray}
 The corresponding creation/annihilation operators are denoted by $a^{\dagger}/a$ and $b^{\dagger}/b$:
\begin{eqnarray}
&&\hat{q}=\frac{1}{\sqrt{2}k}\left(a^{\dagger}+a\right);\;\;\hat{p}=\frac{ik}{\sqrt{2}}\left(a^{\dagger}-a\right),\nonumber\\
&&\hat{q_1}=\frac{1}{\sqrt{2}k'}\left(b^{\dagger}+b\right);\;\;\hat{p_1}=\frac{ik'}{\sqrt{2}}\left(b^{\dagger}-b\right).
\end{eqnarray}
The above transformations are canonical irrespective of the values taken by the factors $k$ and $k'$, which were determined by the restriction that the dangerous terms are vanishing:
\begin{eqnarray}
k&=&\left[\frac{(2I-1)(A_2-A_1)+2jA_1}{(2I-1)(A_3-A_1)+2jA_1}I^2\right]^{1/4},\nonumber\\
k'&=&\left[\frac{(2j-1)(A_2-A_1)+2IA_1+V\frac{2j-1}{j(j+1)}2\sqrt{3}\sin\gamma}{(2j-1)(A_3-A_1)+2IA_1+V\frac{2j-1}{j(j+1)}\sqrt{3}\left(\sqrt{3}\cos\gamma+\sin\gamma\right)}j^2\right]^{1/4}.
\end{eqnarray}
The coupled Hamiltonian has the eigenfunction:
\begin{equation}
|1)_I=\Gamma^{\dagger}_{1}|0)_I,
\end {equation}
where $\Gamma^{\dagger}_{1}$ stands for the phonon operator: 
\begin{equation}
\Gamma^{\dagger}_{1}=X_1a^{\dagger}+X_2b^\dagger-Y_1a-Y_2b,
\label{fonen}
\end{equation}

The phonon amplitudes X and Y are determined such that the phonon defines a boson operator and, at the same time, a harmonic vibration for the quantized form of ${\cal H}$.
The second restriction defines a dispersion equation for the corresponding energy:
\begin{equation}
\Omega^4+B'\Omega^2+C'=0,
\label{ecOm}
\end{equation}
with the coefficients $B'$ and $C'$ having the expressions:
\begin{eqnarray}
B'&=&-\left(\omega_1^2+\omega_2^2+8A_2A_3Ij\right),\nonumber\\
C'&=&\omega_1^2\omega_2^2-4\left(A_3^2k^2k'^2+I^2j^2\frac{A_2^2}{k^2k'^2}\right)\omega_1\omega_2+16A_2^2A_3^2I^2j^2.
\end{eqnarray}
By elementary algebraic manipulation, one finds the $B'=B,\;C'=C$. There exists an interval for variable $\gamma$ where Eq. (2.19) admits two positive solutions,
 ordered as $\Omega_1>\Omega_2$, which define the spectrum of the initial Hamiltonian ${\hat H}$, to be used for describing the experimental data for $^{165,167}$Lu \cite{Scho,Amro}:
\begin{equation}
E_{I,j,n_1,n_2}={\cal H}_{min;I}+\hbar\Omega_{1;I}(n_1+\frac{1}{2})+\hbar\Omega_{2;I}(n_2+\frac{1}{2}).\,\;n_1,\;n_2=0,1,2,....
\label{enerI}
\end{equation}
where ${\cal H}_{min}$ has the expression:
\begin{equation}
{\cal H}_{min}=(A_2+A_3)\frac{I+j}{2}+A_1(I-j)^2-V\frac{2j-1}{j+1}\sin\left(\gamma+\frac{\pi}{6}\right).
\end{equation}
Since each quanta of energy $\Omega_{k,I}$ carries an angular momentum equal to unity which is aligned to the total angular momentum I, the level energies of the three bands are described by:
\begin{eqnarray}
E_{I,j,0,0}&=&{\cal H}_{min;I}+\frac{1}{2}\hbar(\Omega_{1;I}+\Omega_{2;I})\equiv E^{TSD1}_{I},\nonumber\\
E_{I,j,1,0}&=&{\cal H}_{min;I}+\frac{1}{2}\hbar(3\Omega_{1;I}+\Omega_{2;I})\equiv E^{TSD2}_{I+1},\nonumber\\
E_{I,j,2,0}&=&{\cal H}_{min;I}+\frac{1}{2}\hbar(5\Omega_{1;I}+\Omega_{2;I})\equiv E^{TSD3}_{I+2}. 
\label{enTSD}
\end{eqnarray}
Ignoring $j$ and $V$  in the two decoupled terms, one obtains the frequency $\Omega_1=(R-\frac{1}{2})\sqrt{(A_3-A_1)(A_2-A_1)}$, which is the wobbling frequency of the rigid rotor associated to the core.
If ${\bf j}$ keeps the constant value corresponding to the minimum point of ${\cal H}_{min}$, i.e. $j_1=j,\;j_2=j_3=0$, then the system exhibits only one wobbling frequency, namely $\omega_1$.
\section{Comparison with a boson expansion method}
\renewcommand{\theequation}{3.\arabic{equation}}
\setcounter{equation}{0}

About the tensor properties of the bosons $a^{\dagger}, b^{\dagger}$ our comments are as follows. The classical components of the angular momenta are:
\begin{eqnarray}
I^{cl}_{+}&\equiv&\langle \hat{I}_{+} \rangle =\sqrt{r(2I-r)}e^{i\varphi};\;\;I^{cl}_{-}\equiv\langle \hat{I}_{-} \rangle =\sqrt{r(2I-r)}e^{-i\varphi};\;\;I^{cl}_{3}\equiv\langle \hat{I}_{3} \rangle =r-I,\nonumber\\
j^{cl}_{+}&\equiv&\langle \hat{j}_{+} \rangle =\sqrt{t(2j-t)}e^{i\psi};\;\;j^{cl}_{-}\equiv\langle \hat{j}_{-} \rangle =\sqrt{t(2j-t)}e^{-i\psi};\;\;\;j^{cl}_{3}\equiv\langle \hat{j}_{3} \rangle =t-j.
\end{eqnarray}
Passing to the conjugate complex coordinates:
\begin{eqnarray}
C_1&=&\sqrt{2I}\sqrt{\frac{2I-r}{r}}e^{-i\varphi};\;B^{*}_{1}=\frac{1}{\sqrt{2I}}\sqrt{r(2I-r)}e^{i\varphi};\;\rm{with\; the\; Poisson\; bracket}\;\{B^{*}_{1},C_{1}\}=i,\nonumber\\
C_2&=&\sqrt{2j}\sqrt{\frac{2j-t}{t}}e^{-i\psi};\;B^{*}_{2}=\frac{1}{\sqrt{2j}}\sqrt{t(2j-t)}e^{i\psi};\;\rm{with\; the\; Poisson\; bracket}\;\{B^{*}_{2},C_{2}\}=i,\nonumber
\end{eqnarray}
and then quantizing them through the correspondence:
\begin{equation}
(C_{1},B^{*}_{1},\{,\})\to(a,a^{\dagger},-i[,]);\;\;(C_{2},B^{*}_{2},\{,\})\to(b,b^{\dagger},-i[,]),
\end{equation}
one arrives at the Dyson boson \cite{Dys} representation for the angular momenta:
\begin{eqnarray}
\hat{I}^{D}_{+}=\sqrt{2I}a^{\dagger};\;\;\hat{I}^{D}_{-}&=&\sqrt{2I}\left(1-\frac{a^{\dagger}a}{2I}\right)a;\;\;\hat{I}^{D}_{3}=I-a^{\dagger}a,\nonumber\\
\hat{j}^{D}_{+}=\sqrt{2j}b^{\dagger};\;\;\hat{j}^{D}_{-}&=&\sqrt{2j}\left(1-\frac{b^{\dagger}b}{2j}\right)b;\;\;\hat{j}^{D}_{3}=j-b^{\dagger}b.
\end{eqnarray}
From these relations one finds out that $a^{\dagger}$ and $b^{\dagger}$
behave, against rotations, like a tensor of rank one and projection 1. Due to Eq. (\ref{fonen}) this property is also valid for the phonon operator $\Gamma^{\dagger}_{1}$.
This induces also a Dyson boson representation for the odd-system Hamiltonian. Neglecting for a moment the coupling terms, the Hamiltonian becomes a sum of two terms $H_{I}(a^{\dagger},a)$ and 
$H_{j}(b^{\dagger},b)$. To each of these terms, one associates a Schr\"{o}dinger equation by using the Bargmann representation for bosons \cite{Bar}. By a suitable transformation, one separates a potential energy term, which in the harmonic approximation provides the wobbling energies $\omega_1$ and $\omega_2$ (see Ref.\cite{Rad017}). 
To conclude, one may say that within a boson picture our formalism is based on the Dyson boson-representation, which contrasts the description of Ref. \cite{Tan017} where   the Holstein- Primakoff \cite{Hols} boson representation for angular momenta is used. Note that while in Ref.\cite{Tan017} the free term involved in the boson Hamiltonian includes the contributions of higher order boson terms through the normal ordering operation and thus is strongly depending on the expansion truncation, here ${\cal H}_{min}$ is the minimum of the whole energy function. Due to this fact the results of the two approaches are different from each other. In particular, the coefficients $B$ and $C$ from Eq.(\ref{BandC}) are different from $b$ and $c$ of Eq. (22) from \cite{Tan017}. Consequently, the corresponding wobbling energies are different and so are the TSD1, TSD2 and TSD3 level energies. 

\section{Numerical application and discussion}
\renewcommand{\theequation}{4.\arabic{equation}}
\setcounter{equation}{0}
\subsection{Comments on the wobbling motion nature}
\begin{figure}[h!]
\includegraphics[width=0.5\textwidth]{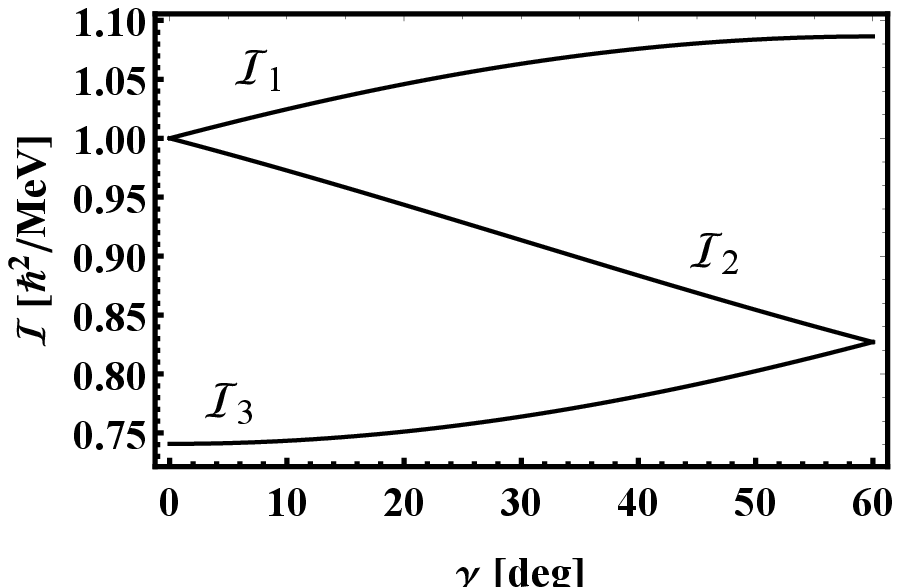}\includegraphics[width=0.5\textwidth]{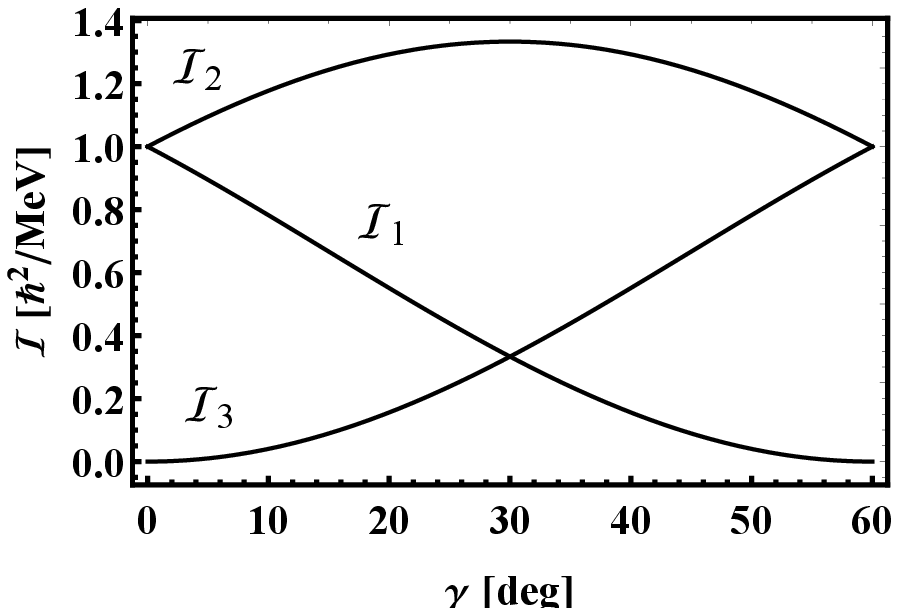}
\begin{minipage}{7.cm}
\caption{The rigid moments of inertia are represented as function of $\gamma$, in units of ${\cal I}_0$.}
\label{Fig. -1}
\end{minipage}\ \
\hspace*{0.5cm}
\begin{minipage}{7.cm}
\caption{The hydrodynamic moments of inertia are plotted as function of $\gamma$, in units of ${\cal I}_0$ . }
\label{Fig.0}
\end{minipage}
\end{figure}

The transverse wobbling mode is the wobbling motion around the middle moment of inertia (MOI). Such a mode does not exist in the
framework  of classical \cite{Land} as well of the  quantum mechanics \cite{BMott}. Coupling a particle of high angular momentum to a triaxial rotor, two major effects show up. The odd particle drives the system to a large deformation, which results in stabilizing the system in a triaxial  strongly deformed shape. Its orientation with respect to the axis of maximum moment of inertia (MOI) may lead either to a longitudinal or to a transverse wobbling, depending whether the particle is aligned to the axis of maximal MOI or is oriented perpendicular to this. The restriction for a transverse wobbling  saying that the middle axis  has the maximal MOI cannot be fulfilled for a rigid MOI, where the maximum MOI corresponds to the first axis, but this might be possible for a hydrodynamic set of MOI, according to Figs. 1 and 2 . Concerning the orientation of the odd particle a.m. this is determined by the particle-core interaction which is attractive for particle-like j, whereas for hole-like j is repulsive. Accordingly, a high j-particle is oriented along the short (s) axis while a high j-hole aligns to the long (l) axis. Indeed, for this configurations the particle-core interaction is minimal. Also, if the odd particle is staying in the middle of the shell, the coupling aligns ${\bf j}$ to the middle axis. For the sake of completeness, in Figs. 1,2 the dependences on $\gamma$ of the rigid and hydrodynamic MOI are shown, respectively. In the Lund convention, i.e. the negative $\gamma$, the hydrodynamic MOI has the expression:
\begin{equation}
{\cal I}^{hyd}_{k}=\frac{4}{3}{\cal I}_0\sin^2(\gamma+\frac{2\pi}{3}k).
\end{equation}
 Also, it is worth mentioning the axes length ordering for the hydrodynamic model:
\begin{equation}
r_1<r_2<r_3 .
\end{equation}
Thus, the 3-axis is the long, the 2-axis the middle, while the 1-axis the short one. In the mentioned convention the triaxiality parameter covers the range $-120^{0}\le\gamma\le 60^{0}$. Within this interval, there are three sectors $[-120^{0},-60^{0}]$, $[-60^{0},0^{0}]$, $[0^{0}, 60^{0}]$ where the system exhibits similar shapes but represents rotations about the long, medium and short axes, respectively. It is common to choose the x axis as rotational axis. Thus, the ordering ${\cal J}_1>{\cal J}_2,{\cal J}_3$ is not fulfilled by the core alone, but the odd particle
whose a.m. is oriented along the x axis,  assists to realize the wobbling.

The specific feature of a transverse wobbling mode is the decreasing behavior of the wobbling frequency, defined as:

\begin{equation}
\hbar\omega_{\omega} = E_{1} (I ) - [E_{0} (I + 1) + E_{0} (I - 1)]/2.
\end{equation}
with increasing the angular momentum, which agrees with the corresponding experimental data. Here, $E_0(I)$ is the energy of the state in the $πi_{13/2}$ band and $E_1 (I )$ is the energy of the state
with spin I in the $n_{\omega} = 1$ wobbling band. Another feature which can be described within  to the transverse mode-formalism, is the large ratio of the inter-band to intra-band E2 transitions, which reflects the collective character of the involved inter-band transition.

 Concluding, the concept of transverse wobbling, proposed by Frauendorf and D\"{o}nau \cite{Frau}, provides a natural
explanation for the decrease of the wobbling frequency with angular momentum and also
for the enhanced E2 transitions between the one phonon wobbling band and the yrast band.

Despite the fact that  the slight decreasing behavior of the wobbling frequency with increasing the angular momentum cannot be described within a longitudinal wobbling regime, there are several publications which  use the mentioned framework and moreover are able to
describe quantitatively both the energies and the transition probabilities \cite{Ham,Ham1,Hage,Tana3,Rad017,Tan017} in all known wobbler's.

As discussed in Ref.\cite{Frau}, in a transverse wobbler there exists a critical angular momentum where a transition from the transverse to longitudinal wobbling takes place. The position of the corresponding state in the wobbling band strongly depends on the mass number, A. Indeed, for $^{163}$Lu this is around 99/2, while for $^{135}$Pr the critical value is about 29/2 \cite{Buda}.
It is interesting to see whether the transverse wobbling persists when the 
frozen alignment of the single particle a.m. hypothesis is relaxed.

It is interesting to mention that in Ref.\cite{Tan017}, Tanabe extended the Holstein-Primakoff boson expansion for a particle-rotor system to a transverse wobbling regime using the hydrodynamic MOI. The application to the case of $^{135}$Pr lead to a negative result: no solution for the wobbling frequency was found. This intriguing result challenges us to extend the present semi-classical formalism to the transverse wobbling. Comparing the results obtained in the two pictures one may decide which is the most appropriate approach for the considered isotopes. Such work is in progress and the results will be published elsewhere. Actually the first step  towards achieving this project is performed in the present paper. 

\subsection{Results for $^{165}$Lu}

Eqs. (\ref{enTSD}) were applied to the case of $^{165}$Lu, where the odd proton is moving in the single j-shell with $j=i_{13/2}$. To the level energies from the three super-deformed bands TSD1,TSD2 and TSD3 the following quantum numbers: $(n_1,n_2)= (0,0);(1,0);(2,0)$ were associated, respectively. It is well established that in the mass region around A=165, the normal-deformed structure coexists with triaxial super-deformed (TSD) shapes \cite{Scho1}. In particular for $^{165}$Lu the total energy surfaces calculated with the Ultimate Cranker code \cite{Beng1,Beng2} exhibits local minima at normal deformation, around $\beta=0.2$ coexisting with a minimum at $\beta=0.38$ with a triaxiality of $\gamma =20^{0}$. These coordinates for the super-deformed minimum are also adopted in the present study.
The formula (\ref{enerI}) involves two parameters, ${\cal I}_0$ and $V$, which were fixed by fitting the experimental excitation energies of the bands TSD1, TSD2 and TSD3, through a least square procedure.  Thus, we obtained for the parameters ${\cal I}_0$ and $V$ the values $52.51568$ MeV/$\rm{\hbar^2}$ and $0.195749$ MeV, respectively.  

The results for the excitation energies characterizing the three super-deformed bands were plotted in Fig.3. From there  we note a good agreement with the data.  

\begin{figure}[ht!]
\includegraphics[width=0.4\textwidth]{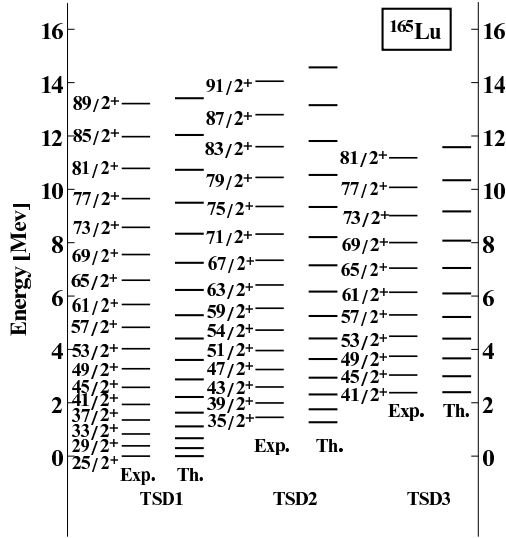}
\caption{Calculated energies for the bands TSD1,TSD2 and TSD3 are compared with the corresponding experimental data taken from Ref.\cite{Scho}.}
\label{Fig.1}
\end{figure}
The validity of the interpretation of  TSD2 and TSD3 as being obtained by exciting with one and two $\Omega_1$-quanta-s is  appraised by comparing the energy differences between the two pairs of bands (TSD1; TSD2) and (TSD1; TSD3) with $\Omega_1$ and 2$\Omega_1$, respectively. This comparison presented in Figs 4 and 5 show that the mentioned hypothesis is quite realistic

\begin{figure}[h!]
\includegraphics[width=0.5\textwidth]{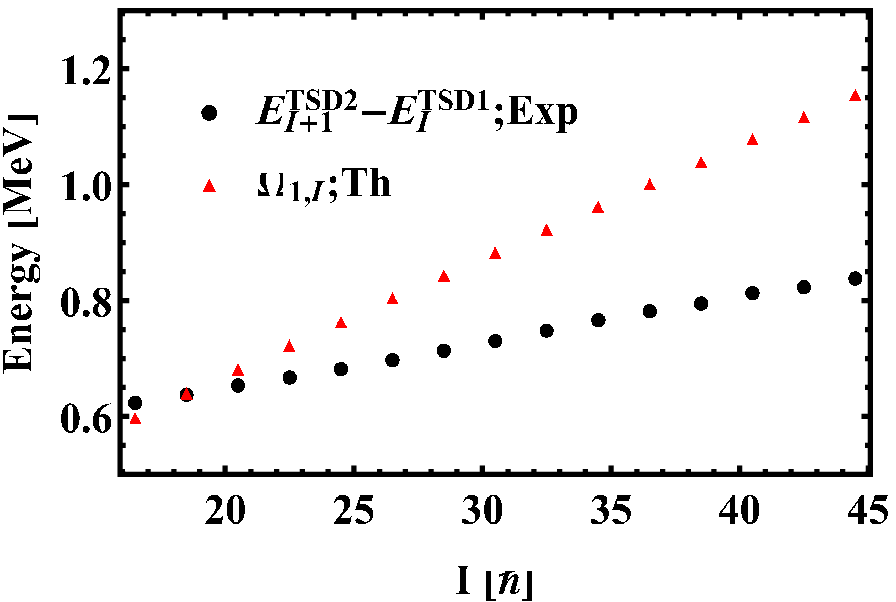}\hspace*{0.1cm}\includegraphics[width=0.5\textwidth]{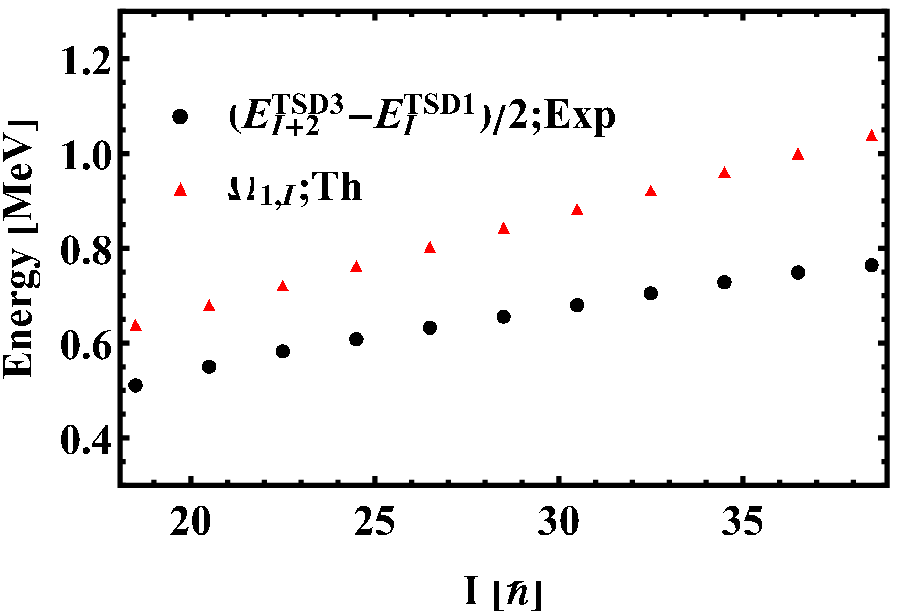}
\begin{minipage}{7.cm}
\caption{The quanta energies $\Omega_{1,I}$ are compared  with the experimental energy differences $E^{TSD2}_{I+1}-E^{TSD1}_{I}$.}
\label{Fig. 2}
\end{minipage}\ \
\hspace*{0.5cm}
\begin{minipage}{7.cm}
\caption{The quanta energies $\Omega_{1,I}$ are compared  with half the experimental energy differences $(E^{TSD3}_{I+2}-E^{TSD1}_{I})/2$. }
\label{Fig.3}
\end{minipage}
\end{figure}

The alignment in the three bands, defined by subtracting from the angular momentum a reference value $I_{ref}={\cal I}_1\omega+{\cal I}_3\omega^3$  with the coefficients ${\cal I}_1$ and ${\cal I}_3$ obtained by a least square procedure fit, is represented in Figs. 6 and 7 for theoretical and experimental data, respectively. Comparing the two figures we note that the theoretical and experimental alignments have a similar behavior as function of the rotational frequency. Moreover, the alignments in the three bands are close to each others. For a large interval of $\omega$, the alignment shows a linear increasing behavior, while for very high frequencies an alignment saturation tendency may be observed, which results in a forward and a slight bending down for the experimental data and results, respectively.

\begin{figure}[h!]
\includegraphics[width=0.5\textwidth]{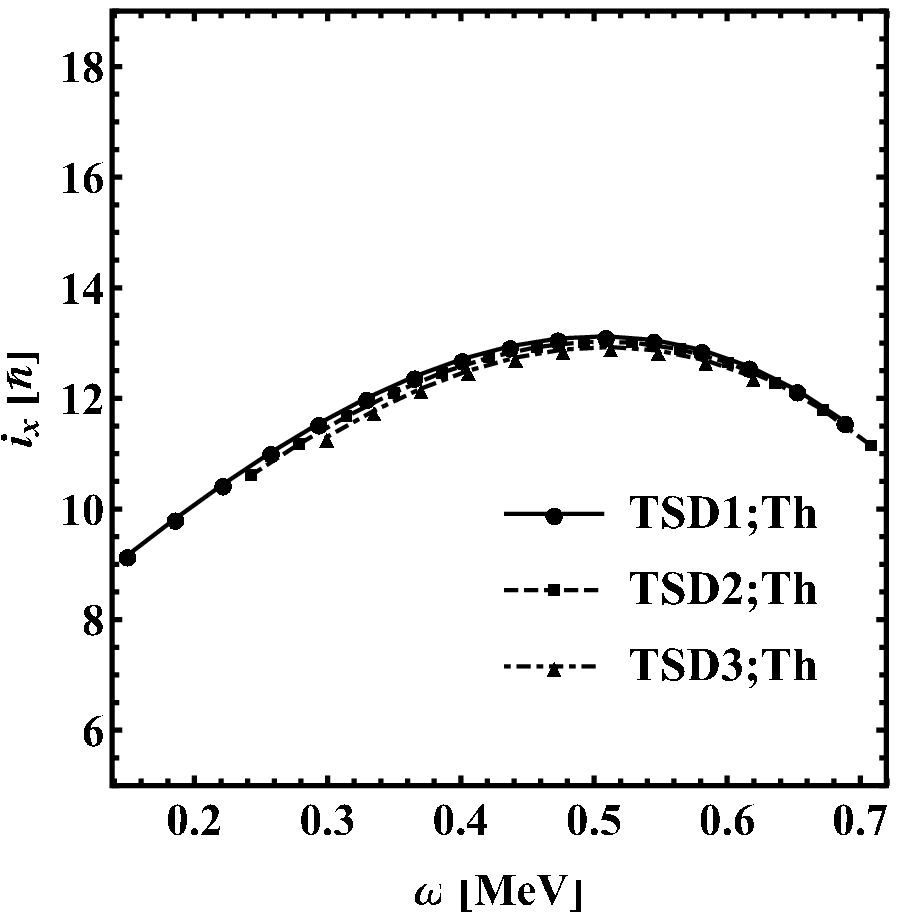}\hspace*{0.2cm}\includegraphics[width=0.5\textwidth]{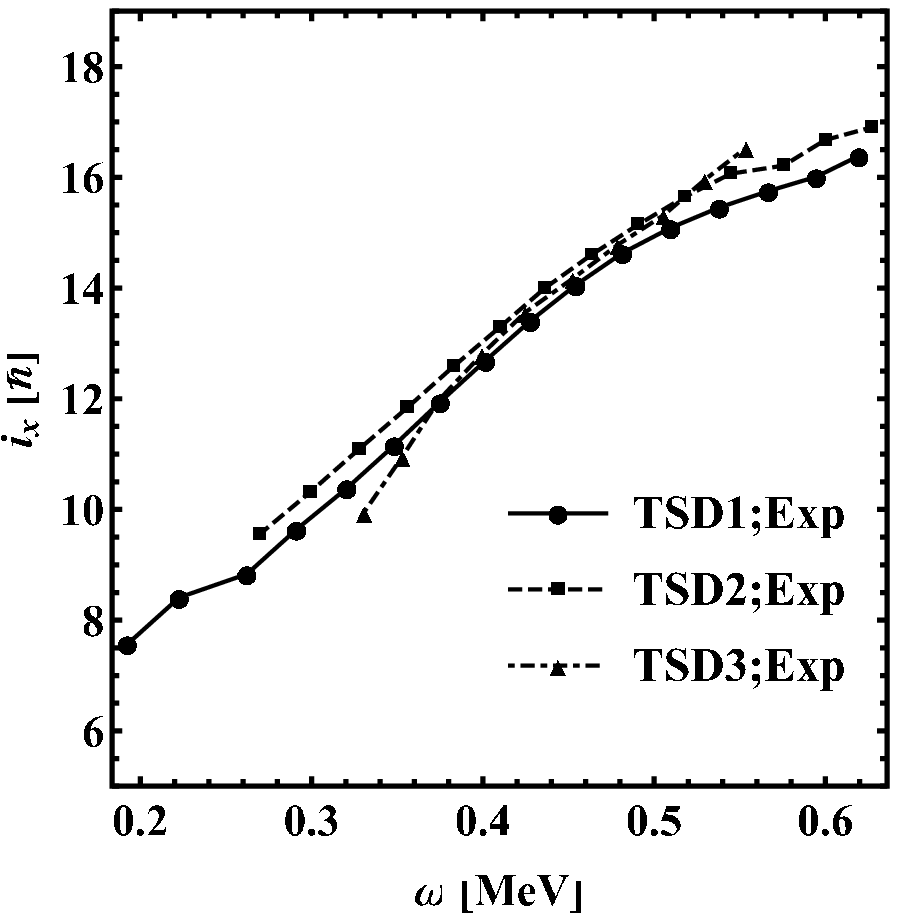}
\begin{minipage}{7.cm}
\caption{The theoretical angular momentum alignment with the reference $I_{ref}={\cal I}_1\omega+{\cal I}_3\omega^3$ with ${\cal I}_1$= 30$\hbar^2$ MeV$^{-1}$ and ${\cal I}_3$= 
40$\hbar^4$MeV$^{-3}$,
is represented as function of the rotational frequency.}
\label{Fig. 4}
\end{minipage}\ \
\hspace*{0.5cm}
\begin{minipage}{7.cm}
\caption{The experimental angular momentum alignment with the reference $I_{ref}={\cal I}_1\omega+{\cal I}_3\omega^3$ with ${\cal I}_1$=30 ${\rm \hbar^2 MeV^{-1}}$ and 
$ {\cal I}_3$=40${\rm\hbar^4MeV^{-3}}$, as a function of the rotational frequency. }
\label{Fig.5}
\end{minipage}
\end{figure}
The dynamic moment of inertia is a  sensitive magnitude at the rotation frequency variation. Its behavior is shown in Fig. 8 for both theoretical and experimental  TSD bands. While in the extreme limits of the $\omega$ interval the experimental dynamic moment of inertia depend on the energy spacings, reflecting an interaction of the states with those from the neighboring normal deformed bands, in the complementary interval this is almost constant for all three bands.
On the other hand, the theoretical dynamic moment of inertia is a constant function of $\omega$ which reflects a linear dependence of $\omega$ on the angular momentum and moreover a similar slope for this dependence in the three bands.
\begin{figure}[t!]
\includegraphics[width=0.5\textwidth]{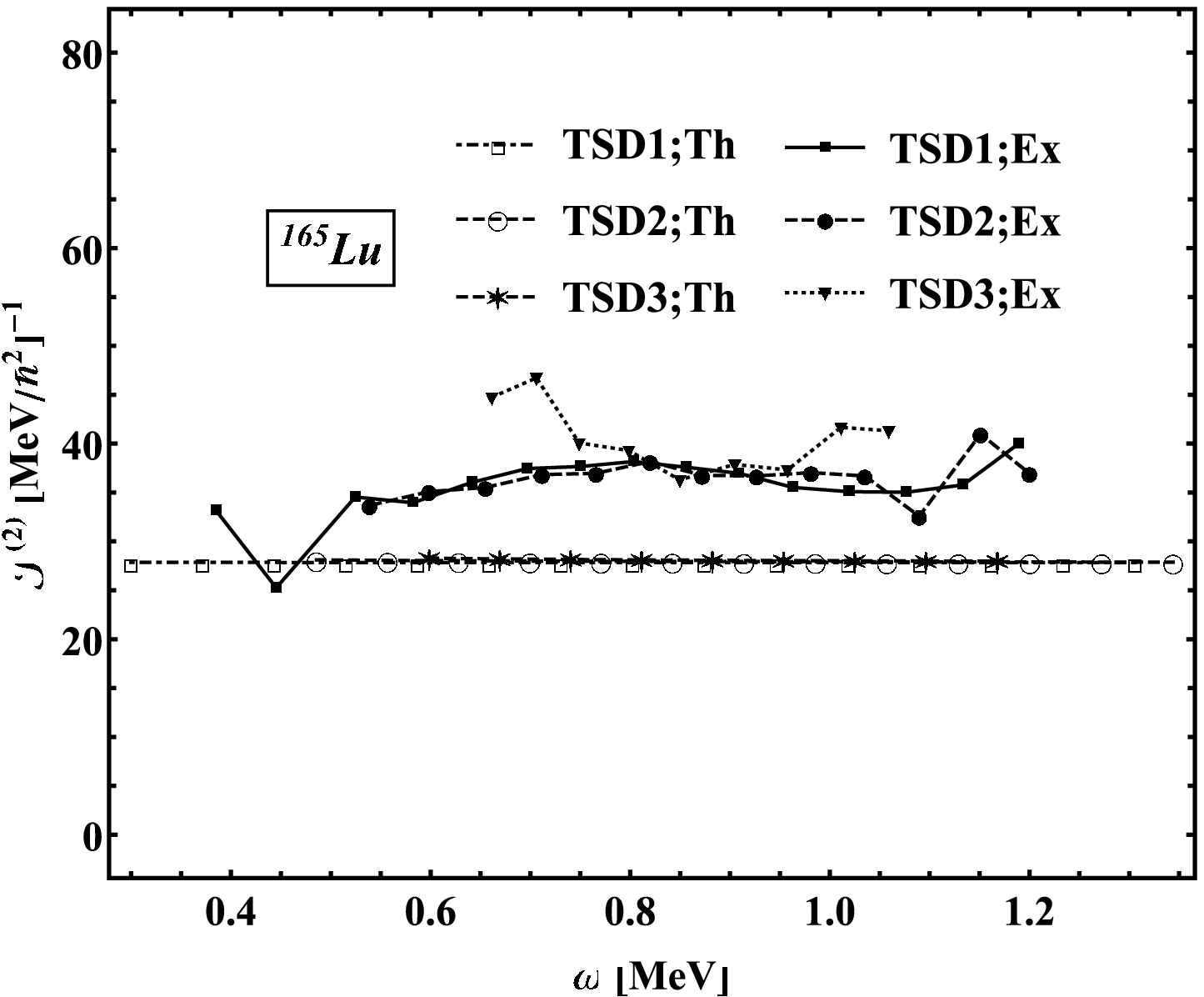}\includegraphics[width=0.5\textwidth]{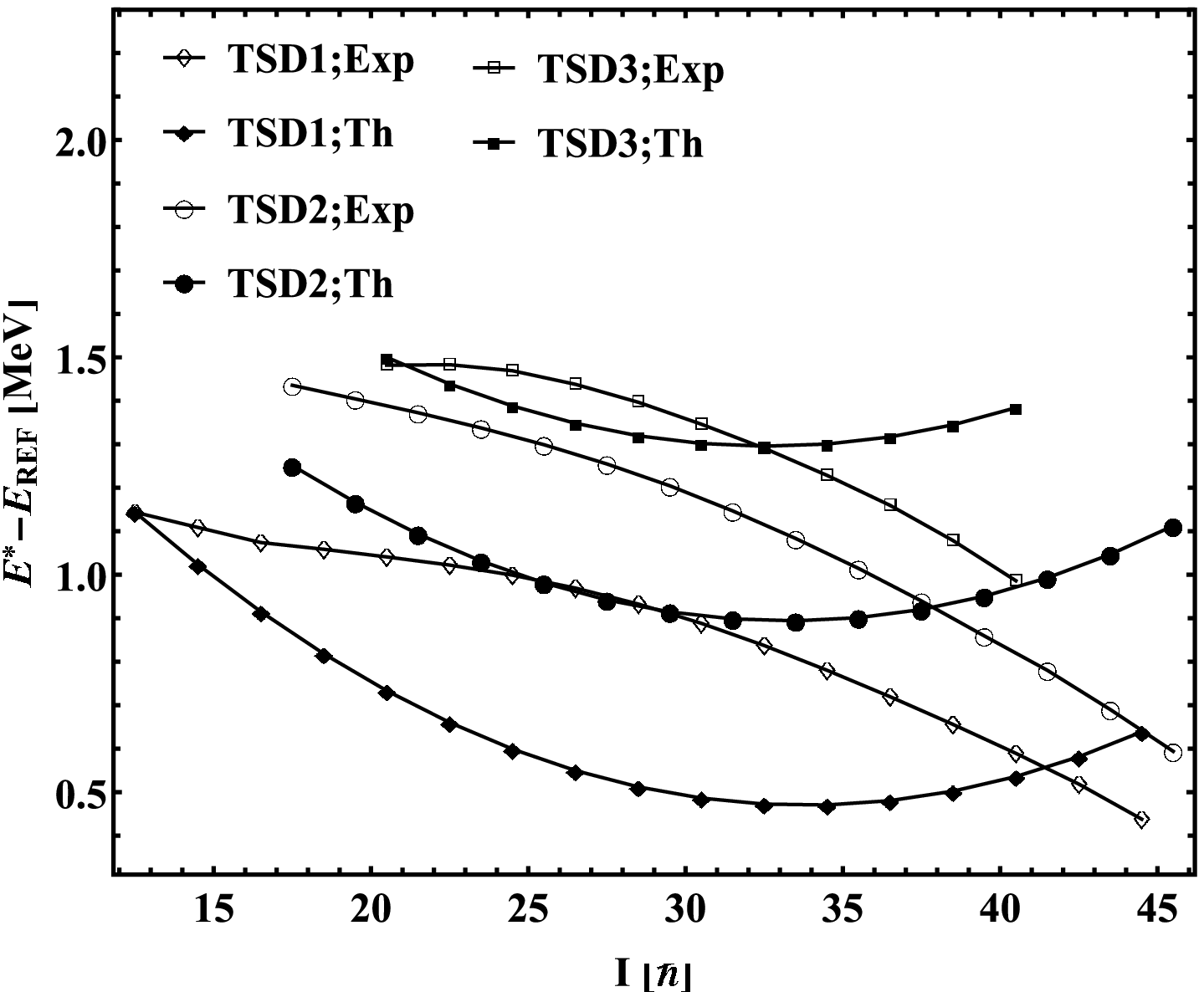}
\begin{minipage}{7.cm}
\caption{The dynamic moment of inertia for TSD1,TSD2 and TSD3, is represented as function of the rotational frequency.}
\label{Fig. 6}
\end{minipage}\ \
\hspace*{0.5cm}
\begin{minipage}{7.cm}
\caption{Theoretical and experimental excitation energies for TSD1,TSD2 and TSD3, normalized to the excitation energy of a rigid rotor with an effective moment of inertia, i.e. 
$E_{REF}=0.0075I(I+1)$[MeV],
are plotted as function of the angular momentum. }
\label{Fig.7}
\end{minipage}
\end{figure}

In Fig.9 the excitation energies of the three bands relative to a rigid rotor reference energy with an effective moment of inertia are plotted as function of angular momentum. Note that the curves
corresponding to the theoretical results are almost parallel and  close to each other. Also, the theoretical and experimental curves for TSD1 and TSD3, intersect twice each other, while those corresponding to TSD2 only once. For a large angular momentum, the departure from the reference energy becomes small, which reflects the alignment effect for large rotation frequency. Inserting the value of the fitted parameter ${\cal I}_0$ in the expression of the rotor coefficients $A_1, A_2$ and $A_3$, the average value of the results define an effective moment of inertia of 
46.63${\rm \hbar^2 MeV^{-1}}$,
which is 1.43 times smaller than the effective moment of inertia of the spherical rigid rotor which is 66.66${\rm \hbar^2 MeV^{-1}}$. Actually, this discrepancy reflects the large effect brought by the triaxial structure of the model Hamiltonian, to the symmetric rigid rotor. The fact that the three bands are characterized by alignments, dynamic moment of inertia and relative energies with respect to an axial rigid rotor with an effective MOI, lying close to each other proves that the three sets of states belong to the same wobbling family.
\begin{figure}[t!]
\hspace*{-0.5cm}\includegraphics[width=0.5\textwidth]{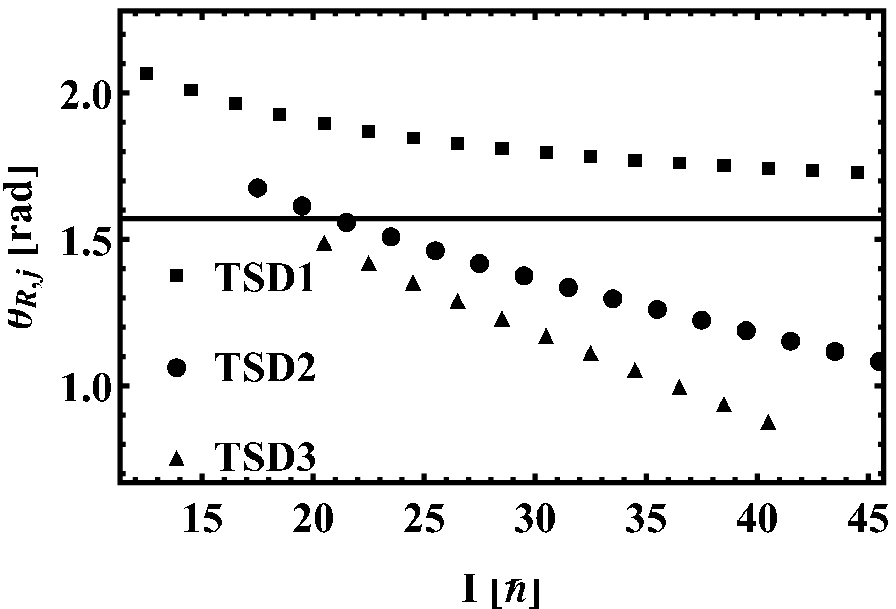}\hspace*{0.1cm}\includegraphics[width=0.5\textwidth]{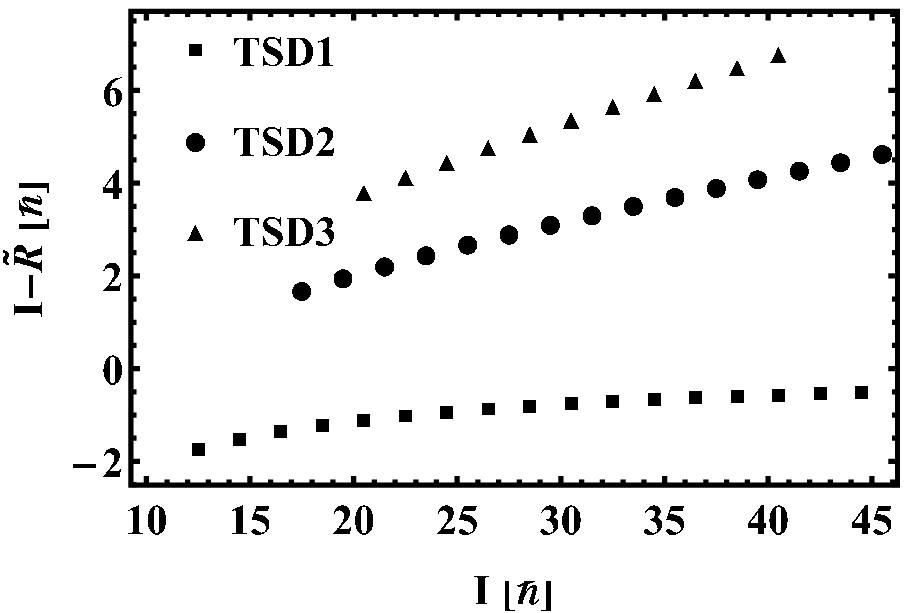}
\begin{minipage}{7.cm}
\caption{The angle between the core, ${\bf R}$, and the single particle, $\bf{j}$, angular momenta in a state from the TSD1 band, as function of the total angular momentum. For comparison, the constant value of $\pi/2$ is also presented.}
\label{Fig. 8}
\end{minipage}\ \
\hspace*{0.5cm}
\begin{minipage}{7.cm}
\caption{The difference $I-\tilde{R}$ as function of I, for the three bands, TSD1,TSD2 and TSD3. }
\label{Fig.9}
\end{minipage}
\end{figure}

It is instructive to calculate the angle between the angular momenta of the core and the odd nucleon.
Within our formalism the mentioned angle is defined as:
\begin{align}
\langle\Phi^{(0)}_{IjM}|\cos\theta|\Phi^{(0)}_{IjM}\rangle =\frac{1}{2}\frac{I(I+1)-\Tilde{R}(\Tilde{R}+1)-j(j+1)}{\sqrt{\Tilde{R}(\Tilde{R}+1)j(j+1)}},
\label{unghi}
\end{align}
 with   
\begin{align}
\Tilde{R}(\Tilde{R}+1)\equiv  \langle\Phi^{(0)}_{IjM}|\hat{R}^2|\Phi^{(0)}_{IjM}\rangle =\sum\limits_{K,\Omega}(\text{C}_{IK})^{2}(\text{C}_{j\Omega})^{2}(\text{C}_{K\ \Omega\ K+\Omega}^{I\ j\ R})^{2}R(R+1); 
\;\;\;
\text{C}_{IK}=\frac{1}{2^{I}}\left(\begin{matrix}2I\cr I-K\end{matrix}\right)^{1/2}.
\end{align} 
Here $|\Phi^{(0)}_{IjM}\rangle$ stands for the wave function $|\Psi_{IjM}\rangle $ (2.5) in the point $(\varphi,r ;\psi, t)$ where the classical energy function reaches its minimum.
From Eq.(\ref{unghi}) the angle $\theta$, which characterizes the band TSD1, can be extracted and the result is shown in Fig. 10. Similarly, one calculates the mentioned angle for TSD2 and TSD3. From Fig. 10 we notice that while for TSD1 the angle goes asymptotically to $\frac{\pi}{2}$,  for the wobbling bands it starts with values close to $\pi/2$ and then  continuously decreases tending to zero for large a.m., which corresponds to the alignment of the particle and the core a.m..
Having $\tilde{R}$ as function of the total a.m. one calculated the difference $I-\tilde{R}$ in order to  see to which extent the core angular momentum aligns to the total angular momentum. This is shown in Fig. 11 where the difference $I-\tilde{R}$ is plotted vs I. From there one sees that for high a.m., in the wobbling bands the core's a.m aligns with I and the plotted difference approaches j (=6.5).Remarkable is the fact that in TSD1 the mentioned difference is negative which reflects the fact that the angle between the a.m. of the core and odd particle is larger than $\frac{\pi}{2}$.

\begin{figure}[ht!]
\includegraphics[width=0.3\textwidth]{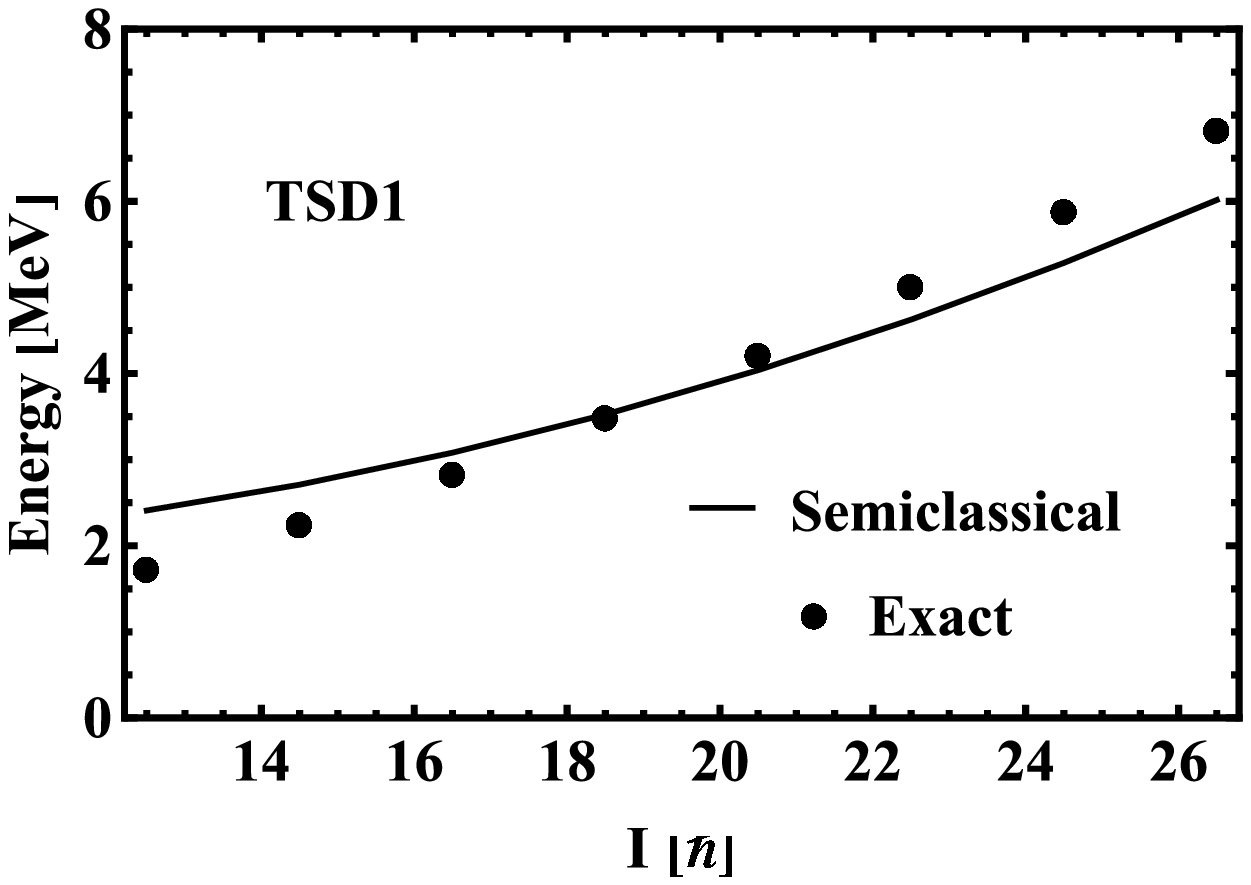}\includegraphics[width=0.3\textwidth]{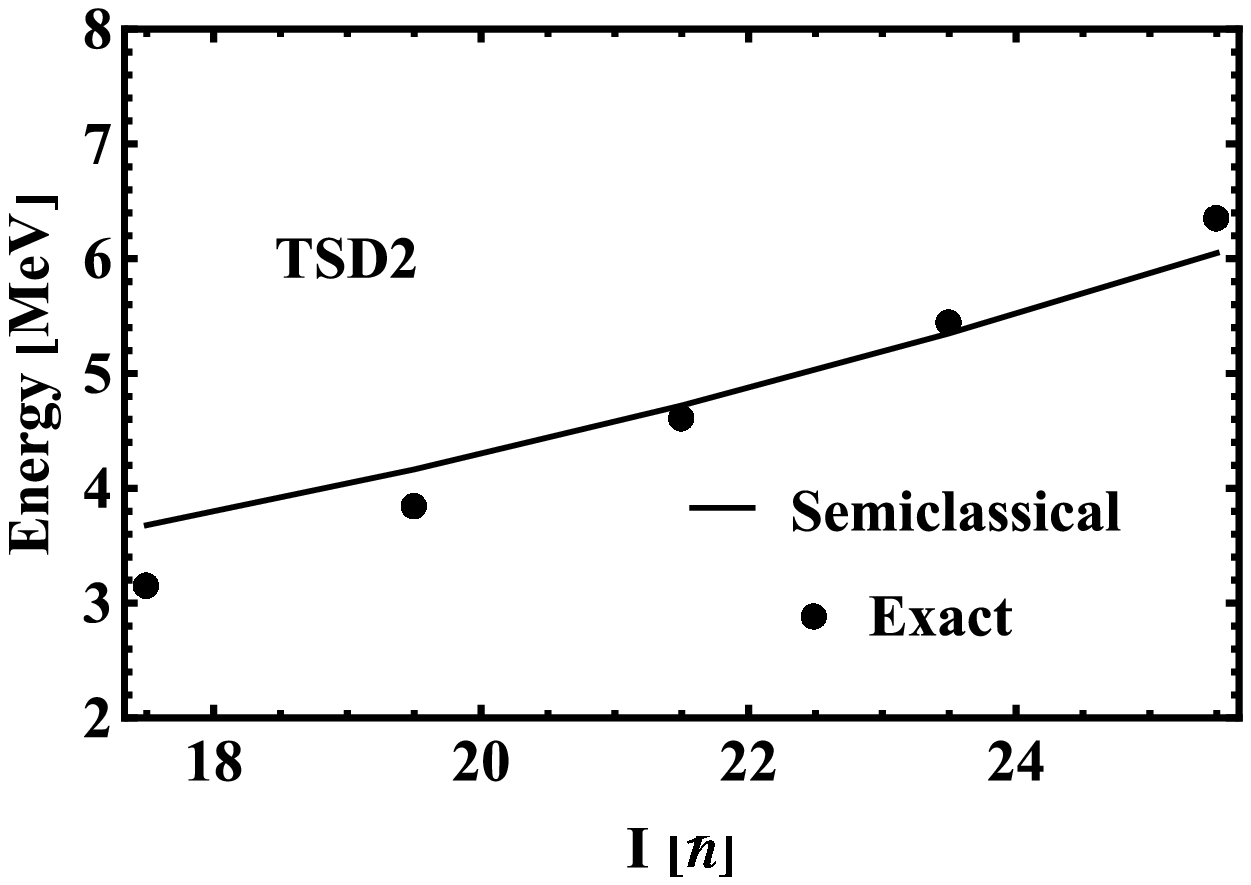}\includegraphics[width=0.3\textwidth]{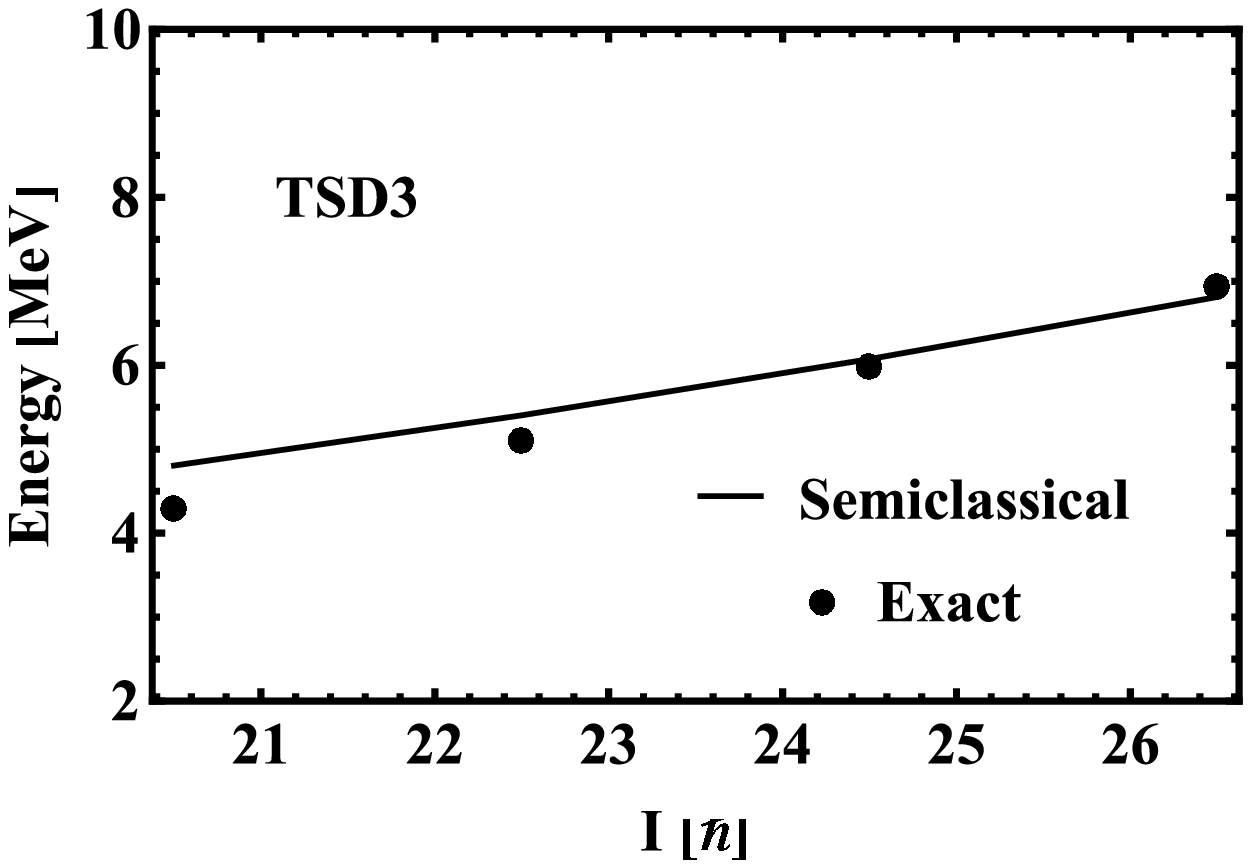}
\caption{Few calculated energies for the bands TSD1, TSD2 and TSD3 are compared with the corresponding exact energies, obtained through the diagonalization of the model particle-core Hamiltonian}
\label{Fig.exact}
\end{figure}

The agreement with experimental data may indicate that some ingredients implied by the single particle motion are already included. However, a few parameters cannot fully cover the complex effects coming from the pairing correlations or the angular momentum dependence of MOI's. Therefore, although the comparison with the experimental data may confirm or not a realistic description, the approximation accuracy information may be obtained by comparing the calculations with the exact results obtained through diagonalization of the model coupling Hamiltonian within the basis:

\begin{equation}
|IKM\rangle=\frac{1}{\sqrt{2j+1}}\sum_{K1,R} C^{R  j  I}_{K_1  \Omega K}|R K_1 M\rangle |j \Omega\rangle .
\end{equation}
For each angular momentum $I$ the resulting eigenvalues are double degenerated, which reflects the presence of the $D_2$ symmetry associated to the core. Considering only one member of the doublet
and denoting by $E_i$ with i=1,2,3,.. the eigenvalues ordered as $E_1>E_2>..$, one notices that the energy spacing is almost constant which, in fact, suggests that the excited states may be obtained by successively applying an energy quanta. Such a structure of the Hamiltonian spectrum has been obtained microscopically in ref.\cite{Raduta84} where the states of even angular momentum  belonging to gamma and beta bands are obtained by exciting the state of the same a.m. from the ground band, with one and two quanta s, respectively. Analyzing the exact energies, another regularity is noticed, namely the energy spacing between the states 13/2+2n and 13/2+2n+1 with $n=1,2,...$, is about the same as the spacing between the state 13/2+2n+1 and the first non-yrast state 13/2+2(n+1). The common energy interval is equal to the wobbling mode energy.  In Fig. \ref{Fig.exact}, few energies from the three bands in
 $^{165}$Lu, are compared with the exact energies. This figure shows a reasonable agreement between the two  sets of data. The noticed deviations might be caused by the adopted assumption of the semi-classical procedure, that the rotor and particle angular momenta are aligned, i.e., $I=R+j$, while in the exact calculations $R$ runs from $|I-j|$ to $I+j$. The advantage of the approach from this paper over the diagonalization method, consists of that the free parameters are easier fixed.

It is worth noting that the fitted moments of inertia are constant along the considered bands, which is not supported by the microscopic calculations. Indeed, the paring interaction diminishes the moment of inertia from the rigid value. Of course, the particle pairs are depaired due to the Coriolis interaction which, is angular momentum dependent. Consequently, the MOI should depend on the angular momentum. Another issue which should be discussed is what happens when several particles are coupled to the core? The rotation tends to align the outer particles angular momenta to the core angular momentum, ending up with an effective angular momentum for the set of coupled particles. The process is gradual, since particles carry different angular momenta and therefore are differently affected by the Coriolis interaction. On the other hand the depairing produces new angular momenta subject to the alignment due to the Coriolis interaction and thus contribute to the increase of the effective angular momentum. One concludes that the one particle-core coupling can be viewed as simulating the coupling with an effective angular momentum accounting for several particles moving around the core.

As mentioned before, the coupling of a high angular momentum particle may drive the system to a large deformation, which
results in stabilizing the system in a triaxial strongly deformed shape. This means that the underlying energy surfaces are relatively soft and due to the coupling the polarization effects play an important role. In particular, the deformation may fluctuate and thus the coupling of the particle to the vibrational modes of the core should be accounted for, which, as a matter of fact, is not the case in the present paper.

Concerning the electric and magnetic reduced transition probabilities only few data are available and these regard the ratios $B(E2)_{out}/B(E2)_{in}$.
The results presented here were obtained with the following wave functions: 

a) a state from the TSD1 band is described by:
\begin{align}
&\Phi^{(1)}_{IjM}={\bf N}_{Ij}\sum_{K,\Omega}C_{IK}C_{J\Omega}|IMK\rangle |j\Omega\rangle\nonumber\\
&\times \left\{1+\frac{i}{\sqrt{2}}\left[\left(\frac{K}{I}k+\frac{I-K}{k}\right)a^{\dagger}+
\left(\frac{\Omega}{j}k'+\frac{j-\Omega}{k'}\right)b^{\dagger}\right]\right\}|0\rangle_{I}.
\end{align}
with ${\bf N}_{Ij}$ standing for the normalization factor and $|0\rangle_{I}$ for the vacuum state of the bosons $a^{\dagger}$ and $b^{\dagger}$. This expression is obtained by the first order expansion of ${\cal H}$ around the minimum point and then quantizing the deviations.

b) the states from the TSD2 band are:
\begin{equation}
|\Phi^{(2)}_{I+1\ j\ M}\rangle = {\cal N}_{I+1}\sum_{M_{1}\mu}C^{I\;1\;I+1}_{M_{1}\mu\;M}|\Phi^{(0)}_{I\; j\;M_{1} }\rangle \Gamma^{\dagger}_{\mu}|0)_{I},\;\;\Gamma^{\dagger}_{\mu}=D^{1}_{\mu\;1}
\Gamma^{\dagger}_{1}.
\end{equation}
The factor ${\cal N}_{I+1}$ makes the function norm equal to unity.
Note that the relation (\ref{unghi}) holds also for the situation when the function $|\Phi^{(0)}_{IjM}\rangle$  is replaced by $|\Phi^{(1)}_{IjM}\rangle$. Actually, such an invariance holds also for the TSD1 energies.

\begin{figure}[ht!]
\includegraphics[width=0.5\textwidth]{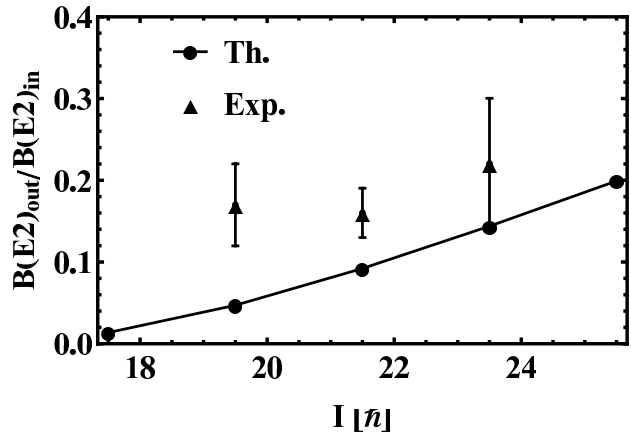}
\caption{The calculated ratios of the BE(2) inter- to intra-band transition, $B(E2)_{out}/B(E2)_{in}$, for few levels from TSD2 are compared with the experimental data taken from Ref. \cite{Scho}. }
\label{Fig.91}
\end{figure}

The transition operators used in the present paper are the same as in Ref.\cite{Rad017}. For an easier reading, these are collected in Appendix C. The results for the ratios $B(E2)_{out}/B(E2)_{in}$ are shown in Fig. 13, where a reasonable agreement with the corresponding experimental data may be seen. The quoted ratio is an increasing function of angular momentum, while in Ref.\cite{Scho} the corresponding function obtained through a cranking formalism, has a different monotony. This behavior of the ratio $B(E2)_{out}/B(E2)_{in}$ as function of the angular momentum is caused by the fact that the inter-band $B(E2)$ values increase faster than those  corresponding to the intra-band transitions. Below the lowest spin shown in Fig. 13, the branching ratio is a decreasing function of angular momentum. It is interesting to notice that the experimental data indicate such a change of monotony at $I=21.5 $.  It is worth mentioning that our approach does not use any adjustable parameter to describe the e.m transition. We considered however an effective charge $e_{eff}=1.5$ which,  as a matter of fact, does not influence the calculated ratio mentioned above.This is at variance, however, with the results reported in Ref. \cite{Rad017} where in order to describe the e.m. transitions a quenching factor of the transition matrix elements for  the TSD2 to TSD1 transitions was necessary. The difference is determined by the fact that in Ref. \cite{Rad017} the excited bands were built up with the low energy phonon, while here the other phonon was used instead. On the other hand it is known that for low energy phonon state  the transition matrix elements  to the ground state are large. Therefore a quenching factor for them, is necessary to be introduced. 

Although there are no other experimental data available, in Tables 1 and 2 we give some additional results for intra- and inter-band transitions. By comparing them with the similar results for $^{163}$Lu we may conclude upon their dependence on the atomic mass. This comparison leads to the conclusion, that the two sets of results are close to each other. The only notable difference regards the transitions quadrupole moments in the band TSD2 which are negative. The difference in sign might be caused by the fact that while in $^{163}$Lu the wobbling phonon is of low energy, in $^{165}$Lu the chosen wobbling phonon is that of larger energy. 
{\scriptsize{
\begin{table}
\begin{tabular}{|c|c|c|c| c| c| c |c|}
\hline
     &                & $B(E2;I^{+}n_w\to(I-2)^+n_{w}')$&$Q_I$& &  & $B(E2;I^{+}n_w\to(I-2)^+n_{w}')$&$Q_I$ \\
     &                &    ${\rm [e^2b^2]}$                          &${\rm [b]}$    & &  &   ${\rm [e^2b^2]}$                            &$ [b]$\\
     &$I^{\pi}$       & $n_w=0;\;\;n_{w}'=0$& &  &$I^{\pi}$       & $n_w=1;\;\;n_{w}'=1$&   \\
\hline
     &$\frac{41}{2}^+$&3.63&10.60&    &$\frac{47}{2}^+$&2.57&-5.66\\
     &$\frac{45}{2}^+$&3.66&10.64&    &$\frac{51}{2}^+$&2.67&-5.76\\
     &$\frac{49}{2}^+$&3.69&10.68&    &$\frac{55}{2}^+$&2.76&-5.84\\
TSD1 &$\frac{53}{2}^+$&3.71&10.71&TSD2&$\frac{59}{2}^+$&2.83&-5.91\\
     &$\frac{57}{2}^+$&3.73&10.73&    &$\frac{63}{2}^+$&2.90&-5.98\\
     &$\frac{61}{2}^+$&3.75&10.76&    &$\frac{67}{2}^+$&2.96&-6.03\\
     &$\frac{65}{2}^+$&3.77&10.78&    &$\frac{71}{2}^+$&3.01&-6.08\\
     &$\frac{69}{2}^+$&3.78&10.79&    &                &    &     \\
\hline
\end{tabular}
\caption{Intra-band B(E2) values and transition quadrupole moments for TSD1 an TSD2. As expected, for TSD1 the $Q_I$ values are large.}
\end{table}}}
{\scriptsize{
\begin{table}
\begin{tabular}{|c|c|c|c|}
\hline
                     & $B(E2;I^{+}n_w\to(I-1)^+n_{w}')$& $B(M1;I^{+}n_w\to(I-1)^+n_{w}')$&$\delta_{I\to (I-1)}$ \\
                     &    ${\rm [e^2b^2]}$                          &${\rm [\mu_{N}^{2}]}$    & ${\rm [MeV.fm]}$\\
     $I^{\pi}$       & $n_w=1;\;\;n_{w}'=0$           & $n_w=1;\;\;n_{w}'=0$& $n_w=1;\;\;n_{w}'=0$  \\
\hline
$\frac{47}{2}^+$&0.37&0.148&-0.958\\
$\frac{51}{2}^+$&0.53&0.176&-1.054\\
$\frac{55}{2}^+$&0.71&0.205&-1.251\\
$\frac{59}{2}^+$&0.89&0.235&-1.372\\
$\frac{63}{2}^+$&1.07&0.267&-1.483\\
\hline
\end{tabular}
\caption{The B(E2) and B(M1) values for the transitions from TSD2 to TSD1. The corresponding mixing ratios are also mentioned.}
\end{table}}}

The proposed approach is pure phenomenological, developed on the base of minimum action Principle and  has no  microscopic counterpart.

\subsection{The case of $^{167}$Lu}
Here we describe the results of our calculation for $^{167}$Lu, although this is almost isospectral with $^{165}$Lu. Results are collected in six panels of Fig.14 and one table, table III.
The odd particle is moving in a single j shell which is the intruder $i_{13/2}$. The core is a triaxial rotor with MOI given, in the Lund convention, by Eq. (2.3). 
The nuclear quadrupole deformation $\beta$ and the departure from the symmetric symmetry $\gamma$, were taken as suggested by the Ultimate Cranker (UC) calculation of the potential energy surface.
Indeed, the mentioned potential exhibits two minima, one  normal deformed minimum and a super-deformed minimum, with $\beta=0.4$ and $\gamma =20^0$. Thus, the three MOI parameters are fully determined by ${\cal I}_0$ and $V$, which were fixed by fitting the experimental excitation energies in the two bands, through the least square procedure. The fixed values are 
${\cal I}_0=52.6287 {\rm [\hbar^2MeV^{-1}]}$ and $V=1.01825{\rm MeV}$. The calculated energies are represented as function of a.m. in the upper row for TSD1 (the left panel) and TSD2 (right panel). In the two panels of the second row the theoretical alignment is compared with the experimental one. In both cases the alignment is an increasing function for small rotational frequency and a decreasing function for large frequency.
In the third row in Fig. 14, the left panel, one gives the energy relative to a reference energy characterizing a symmetric rigid rotor with an effective MOI.The curves for TSD1 and TSD2 have a similar behavior , as function of the total a.m., I. The theoretical curves intersect the experimental ones at low and high a.m.. Having ${\cal I}_0$ fixed one can determined the individual MOI parameters:
${\cal I}_1=55.6257{\rm \hbar^2MeV^{-1}}$,\;\;   ${\cal I}_2=48.9524 {\rm\hbar^2MeV^{-1}}$ ,\;\;  ${\cal I}_1=36.4108 {\rm\hbar^2MeV^{-1}}$ to which corresponds an average MOI of about 
$47 {\rm\hbar^2 MeV^{-1}}$.
This is to be compared with the value of $66.66 {\rm\hbar^2MeV^{-1}}$, which determines the reference energy from the above mentioned figure. Certainly, the difference is caused by the departure from the axial symmetric picture.

The dynamic moment of inertia is plotted in the last panel of the last row, as function of the rotation frequency. In the experimental plot four point depart from the average values by a consistent amount.
These deviations are determined by the interaction of the corresponding states of TSD1 with the neighboring states of a normal deformed band. This is also reflected in the alignment picture where
the first two points are lowered and moreover a kink which shows up just where the staggering of the dynamic moment of inertia manifests. The interaction of TSD1 band and the neighboring ND band is not considered in our formalism and therefore the  dynamic moments of inertia in the two bands  almost coincide and moreover are constant. This suggests that the rotation frequency is a constant function of I.

\begin{figure}[h!]
\includegraphics[width=0.4\textwidth]{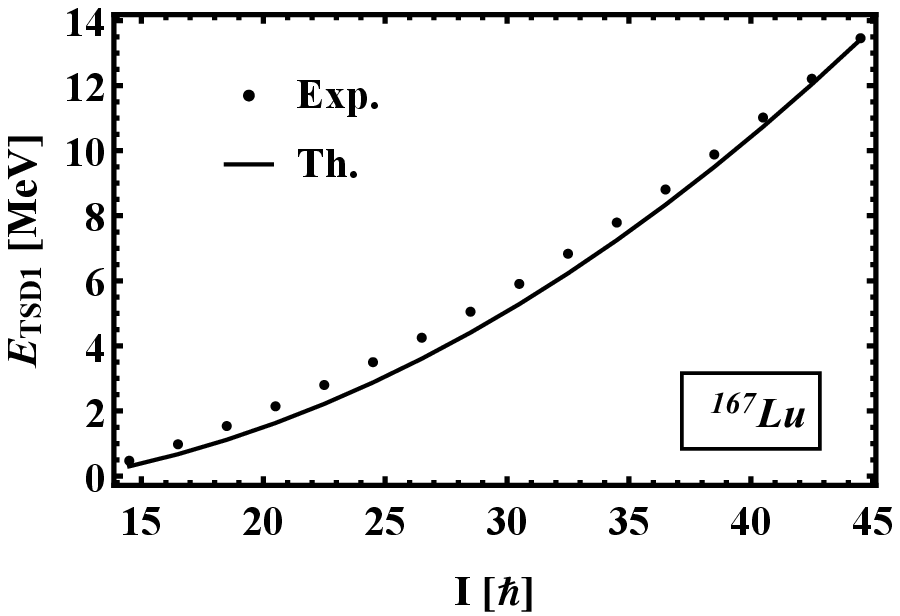}\includegraphics[width=0.4\textwidth]{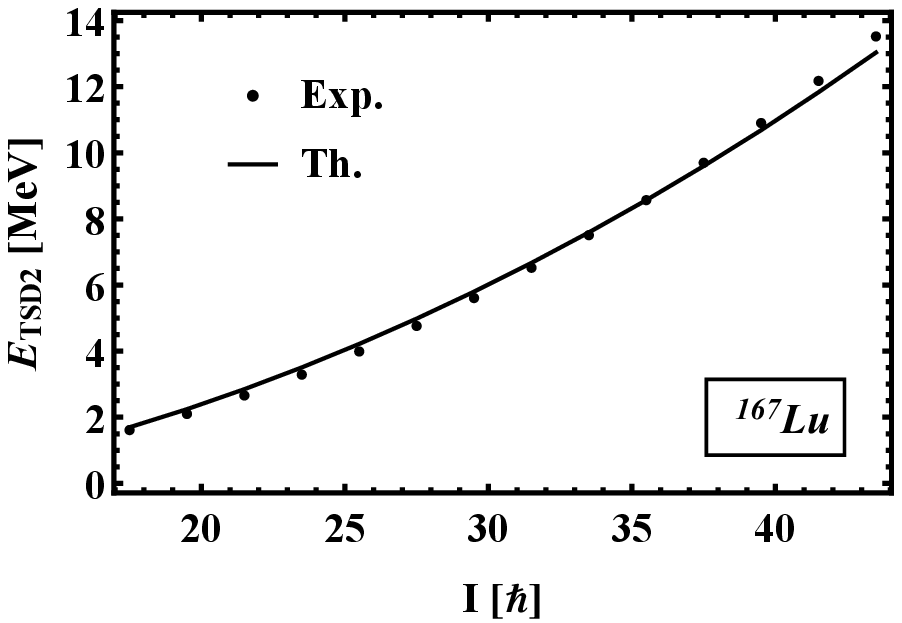}
\includegraphics[width=0.4\textwidth]{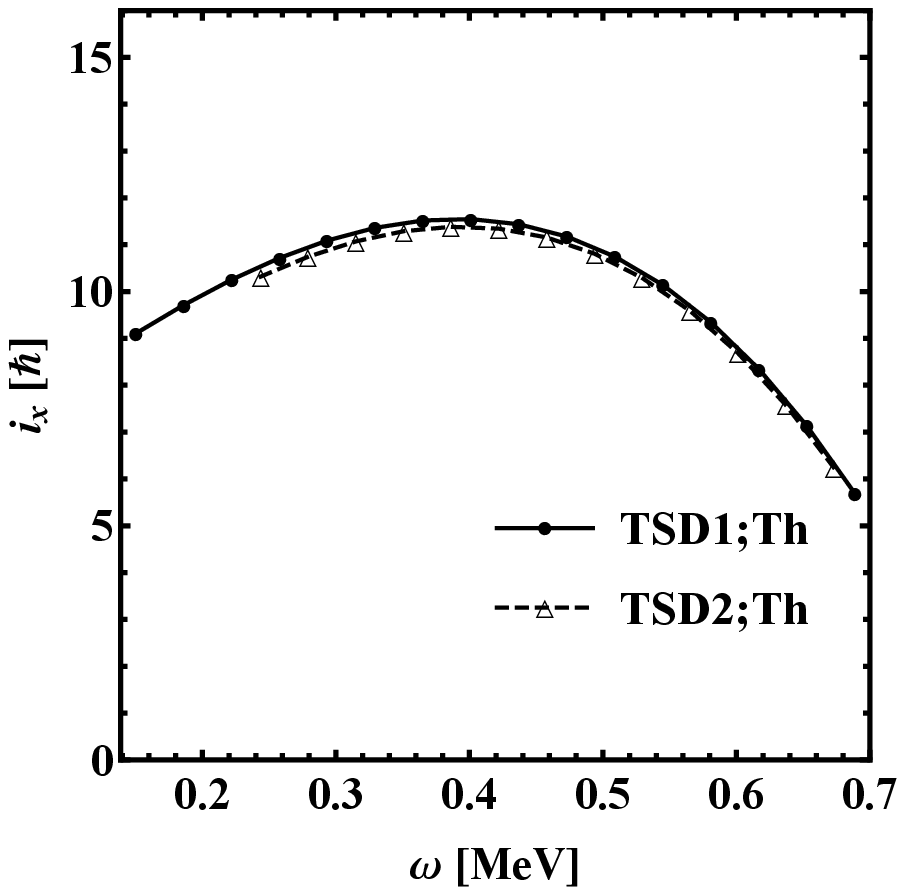}\includegraphics[width=0.4\textwidth]{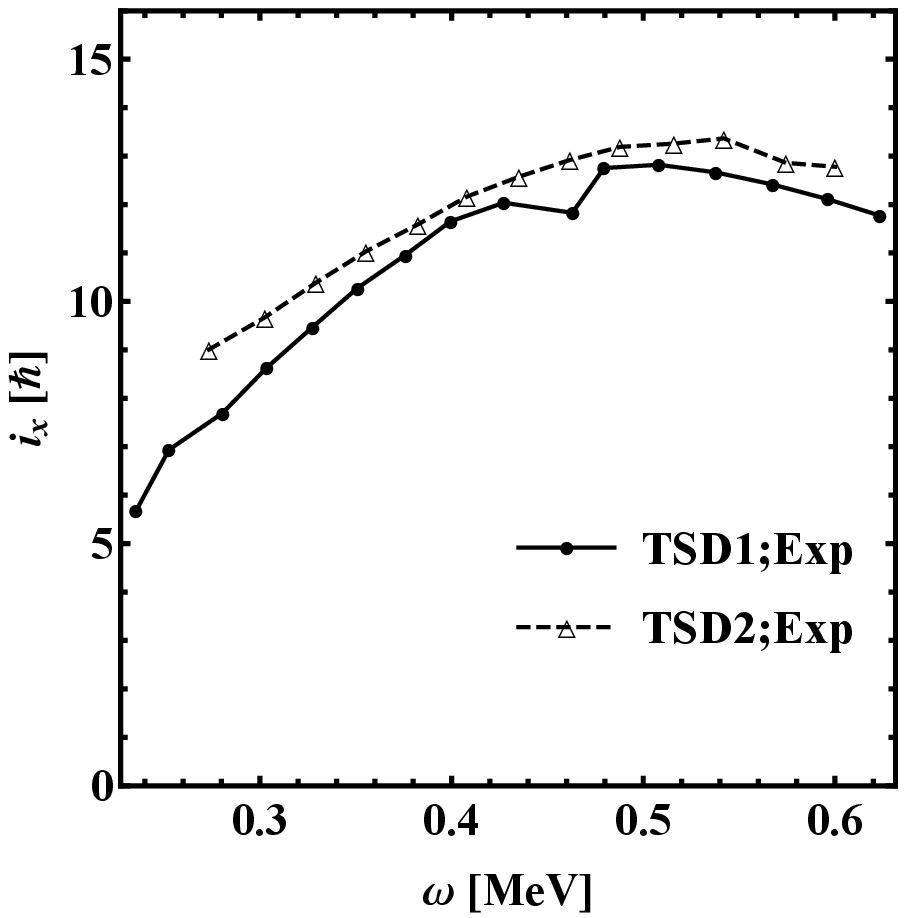}
\includegraphics[width=0.4\textwidth]{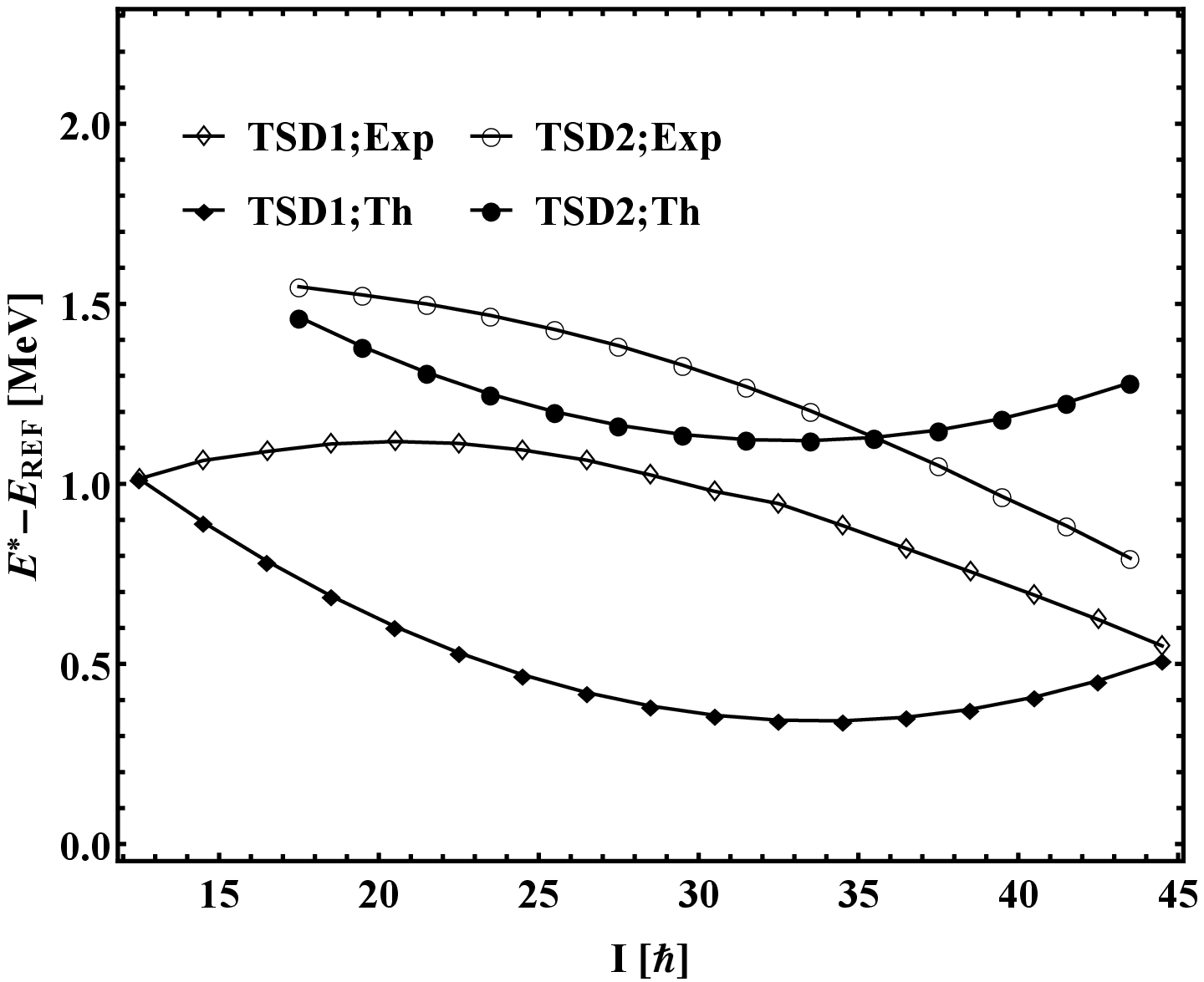}\includegraphics[width=0.4\textwidth]{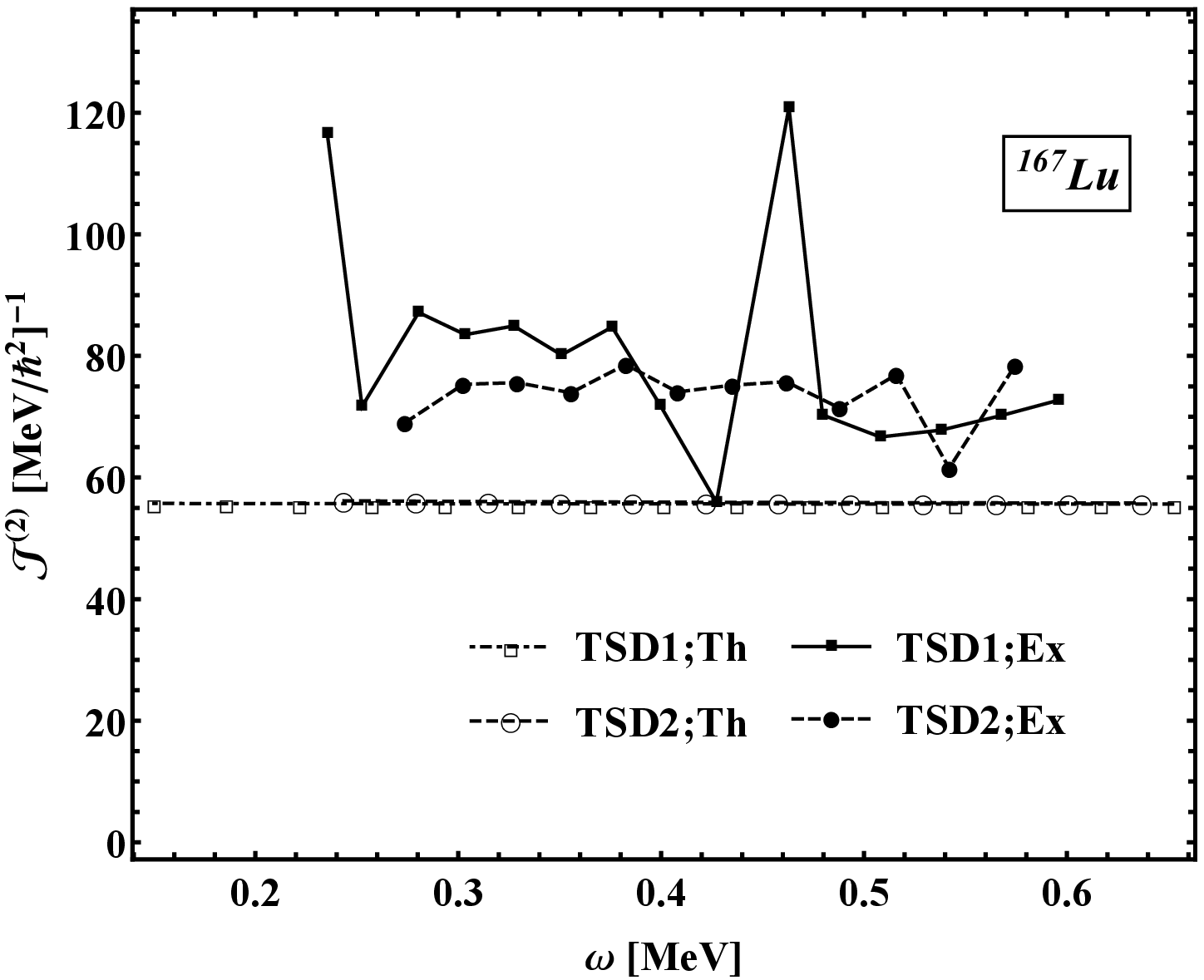}
\caption{First row: excitation energies for the bands TSD1 (left) and TSD2(right) given in MeV;
         second row:the calculated(left) and experimental alignment(right) for $^{167}$Lu. The reference a.m. is 
$I^{ref}={\cal I}_1\omega+{\cal I}_3 \omega^3$ with ${\cal I}_1=35{\rm \hbar^2 MeV^{-1}}$  and ${\cal I}_3=45 {\rm\hbar^4 MeV^{-3}}$;
third row: The relative energies (left) for the TSD1 and TSD2 bands with respect to a reference energy characterizing a rigid axial symmetric rotor, i.e, $E_{ref}=0.0075I(I+1){\rm[MeV]}$;
right: the calculated dynamic moment of inertia of states from TSD1 and TSD2, are compared with the corresponding experimental data.}
\end{figure}

\begin{table}[h!]
\begin{tabular}{|c|cc|cc|}
\hline
&\multicolumn{2}{c|}{$\frac{B(E2)_{out}}{B(E2)_{in}}$}&\multicolumn{2}{c|}{$\delta_{I\to(I-1)}[MeV.fm]$}\\
$I^{\pi}$&Th.  & Exp.&Th.&Exp.\\
\hline
$\frac{39}{2}^+$& 0.72       &$0.23^{+0.02}_{-0.05}$&  -6.45        &$-3.1^{+1.1}_{-3.4}$\\
$\frac{43}{2}^+$& 0.57       &    -                &  -5.37        &- \\
$\frac{47}{2}^+$& 0.45       &$0.26^{+0.03}_{-0.04}$&  -4.51        &$-5.1^{-1.6}_{-2.5}$\\
$\frac{51}{2}^+$& 0.35       &$0.27^{+0.07}_{-0.10}$&   -3.81       &$ -3.9^{+2.7}_{-8.4}$\\
$\frac{55}{2}^+$& 0.28      &     -               &   -3.23       & -\\
\hline
\end{tabular}
\caption{the ratios of the E2 inter-band to intra-band transitions for the TSD2 band. Also the mixing ratios are given. Results of our calculations are compared with the corresponding experimental data.}
\end{table}

\section{conclusions}
\renewcommand{\theequation}{5.\arabic{equation}}
\setcounter{equation}{0}
Summarizing, various features of the wobbling motion in $^{165,167}$Lu were interpreted within a semi-classical formalism. Thus, the particle-triaxial core Hamiltonian was dequantized via the minimum action Principle which provides a set of Hamilton equations for the classical phase space coordinates, which are coherence parameters of a coherent state for the group $SU(2)\otimes SU(2)$ generated by the intrinsic components of the total angular momentum and of the angular momentum for the odd nucleon, respectively.  The equations of motion are first brought to a canonical form and then
linearized by expanding them around the minimum of the classical energy. The solutions of the linearized equations define two harmonic vibrations whose energies added to the minimum classical energy represent analytical formulas for the energy levels of the experimental bands, TSD1, TSD2 and TSD3 in $^{165}$Lu  and TSD1, TSD2 in $^{167}$Lu. The parameters involved are determined by a least square procedure. 
Quantizing the classical coordinates, the trial function and the first phonon excitation of it become model states for the bands TSD1 and TSD2. Furthermore, they are used to calculate the reduced electric and magnetic transition probabilities.  The departure from the aligned picture and the rigid rotor with and effective moment of inertia are also analyzed. Also the dynamic moment of inertia and its a.m. dependence are calculated. From these issues we learn that in each nucleus the considered bands have similar properties which in fact reflects their affiliation to the same wobbling family. 
The angle between the core and single particle a.m is calculated for the states of the three bands in $^{165}$Lu. The difference between the total and the core a.m. is interpreted as their alignment. Also, for the wobbling bands and large a.m., the odd nucleon aligns its angular momentum to the core's angular momentum, due to the Coriolis interaction. The electric and magnetic transition probabilities were calculated and compared with the available data. One remarks on a good agreement with the data in both nuclei considered in the present paper.

A full subsection is devoted to summarizing the 
main signatures of the transverse wobbling which is mainly based on the frozen alignment hypothesis and the assumption that the odd particle a.m. is oriented perpendicular to the axis of maximum moment of inertia. Actually this allows a decreasing behavior of the wobbling frequency on the total a.m. This feature is not accounted for by any previous phenomenological models although all the other properties are quantitatively described. Moreover in Ref.\cite{Tan017} the extension of the Holstein-Primakoff boson expansion to the transverse wobbling regime provided an equation for the wobbling frequency which has no real solutions. Therefore more investigations are needed in order to bring the formalism proposed by Frauendorf and Donau [32]and  other 
approaches [9,11,13,14,21,26] in agreement with each other. In this context the extension of the present semi-classical description to the case of a transverse wobbling is in progress and the results will be published elsewhere.

Concluding, the present results  prove that the proposed semi-classical model is a suitable tool to account for various features of the wobbling motion in $^{165,167}$Lu, in a realistic fashion.
\section{Appendix A}
\renewcommand{\theequation}{A.\arabic{equation}}
\setcounter{equation}{0}
The equations of motion provided by the variational principle are:
\begin{eqnarray}
\stackrel{\bullet}{\varphi}&=&\frac{2I-1}{I}(I-r)\left(A_1\cos^2\varphi+A_2\sin^2\varphi-A_3\right)\nonumber\\
                           &-&2\sqrt{\frac{t(2j-t)}{r(2I-r)}}(I-r)\left(A_1\cos\varphi\cos\psi+A_2\sin\varphi\sin\psi\right)+2A_3(j-t),\nonumber\\
\stackrel{\bullet}{\psi}&=&\frac{2j-1}{j}(j-t)\left(A_1\cos^2\psi+A_2\sin^2\psi-A_3\right)\nonumber\\
                           &-&2\sqrt{\frac{r(2I-r)}{t(2j-t)}}(j-t)\left(A_1\cos\varphi\cos\psi+A_2\sin\varphi\sin\psi\right)+2A_3(I-r)\nonumber\\
                           &-&V\frac{2j-1}{j^2(j+1)}(j-t)\sqrt{3}\left(\sqrt{3}\cos\gamma+\sin\gamma\cos2\psi\right),\nonumber\\
-\stackrel{\bullet}{r}&=&\frac{2I-1}{2I}r(2I-r)\left(A_2-A_1\right)\sin2\varphi\nonumber\\
                           &+&2\sqrt{t(2j-t)r(2I-r)}\left(A_1\sin\varphi\cos\psi-A_2\cos\varphi\sin\psi\right),\nonumber\\
-\stackrel{\bullet}{t}&=&\frac{2j-1}{2j}t(2j-t)\left(A_2-A_1\right)\sin2\psi\nonumber\\
                      &+&2\sqrt{r(2I-r)t(2j-t)}\left(A_1\cos\varphi\sin\psi-A_2\sin\varphi\cos\psi\right)\nonumber\\
                           &+&V\frac{2j-1}{j^2(j+1)}t(2j-t)\sqrt{3}\sin\gamma\sin2\psi .
\end{eqnarray}

From these equations one readily finds that the function ${\cal H}$ is a constant of motion, i.e. $\stackrel{\bullet}{\cal H}=0$. This equation defines a surface,  called as equi-energy surface, ${\cal H}=const.$ Actually this is a consequence of the fact that the equations of motion are derived from a variational principle. Also, one notes that the stationary coordinates, having vanishing time derivatives, are stationary points for the equi-energy surface. There are several stationary points, among which some are minima as suggested by the sign of the associated Hessian. For example, one minimum is achieved in the point
$(r,\varphi;t, \psi)=(I,0;j,0)$

Expanding the classical energy function around the minimum point and denoting the deviations from the minimum by prime letters one obtains:
\begin{eqnarray}
{\cal H}&=&{\cal H}_{min}+\frac{1}{I}\left((2I-1)(A_3-A_1)+2jA_1\right)\frac{r^{\prime 2}}{2}+I\left((2I-1)(A_2-A_1)+2jA_1\right)\frac{\varphi^{\prime 2}}{2}\nonumber\\
        &+&\frac{1}{j}\left[(2j-1)(A_3-A_1)+2IA_1+V\frac{2j-1}{j(j+1)}\sqrt{3}\left(\sqrt{3}\cos\gamma+\sin\gamma\right)\right]\frac{t^{\prime 2}}{2}\nonumber\\
        &+&j\left[(2j-1)(A_2-A_1)+2IA_1+V\frac{2j-1}{j(j+1)}2\sqrt{3}\sin\gamma \right]\frac{\psi^{\prime 2}}{2}\nonumber\\
        &-&2A_3r^{\prime}t^{\prime}-2IjA_2\varphi^{\prime}\psi^{\prime}.
\end{eqnarray}
Neglecting,for the moment, the coupling terms one readily obtains that the classical energy is the sum of two independent oscillators whose frequencies are:
\begin{eqnarray}
\omega_1&=&\left[\left((2I-1)(A_3-A_1)+2jA_1\right)\left((2I-1)(A_2-A_1)+2jA_1\right)\right]^{1/2},\nonumber\\
\omega_2&=&\left[(2j-1)(A_3-A_1)+2IA_1+V\frac{2j-1}{j(j+1)}\sqrt{3}\left(\sqrt{3}\cos\gamma+\sin\gamma\right)\right]^{1/2}\nonumber\\
&\times&\left[(2j-1)(A_2-A_1)+2IA_1+V\frac{2j-1}{j(j+1)}2\sqrt{3}\sin\gamma \right]^{1/2}.
\end{eqnarray}
In order to have real solutions for the two frequencies, the MOI parameters and the single particle potential strength V must fulfill some restrictions:
\begin{eqnarray}
&&S_{Ij}A_1<A_2<A_3,\; {\rm or}\; \;S_{Ij}A_1<A_3<A_2,\nonumber\\
&&A_3>T_{Ij}A_1+V_1,\;\;A_2>T_{Ij}A_1+V_2.
\end{eqnarray}
where the following notations have been used:
\begin{eqnarray}
&&S_{Ij}=\frac{2I-1-2j}{2I-1},\;\;T_{Ij}=\frac{2j-1-2I}{2j-1},\nonumber\\
&&V_1=\frac{V}{j(j+1)}\sqrt{3}(\sqrt{3}\cos\gamma+\sin\gamma ),\;\;V_2=\frac{V}{j(j+1)}2\sqrt{3}\sin\gamma .
\end{eqnarray}

To treat the coupling term involved in the energy function it is useful to  quantize the phase space coordinates:
\begin{eqnarray}
&&\varphi\to\hat{q};\;\;r\to \hat{p};\;\; [\hat{q},\hat{p}]=i,\nonumber\\
&&\psi\to\hat{q}_1;\;\;t\to \hat{p}_1;\;\; [\hat{q}_1,\hat{p}_1]=i.
\end{eqnarray}
We associate to the two oscillations, defined above, the creation/annihilation operators:
\begin{eqnarray}
&&\hat{q}=\frac{1}{\sqrt{2}k}(a^{\dagger}+a),\;\;\hat{p}=\frac{ik}{\sqrt{2}}(a^{\dagger}-a),\nonumber\\
&&\hat{q}_1=\frac{1}{\sqrt{2}k'}(b^{\dagger}+b),\;\;\hat{p}_1=\frac{ik'}{\sqrt{2}}(b^{\dagger}-b).
\end{eqnarray}
The transformation relating the coordinates and the conjugate momenta with the operators $a^{\dagger}, a$ and $b^{\dagger}, b$ is canonical irrespective the value of the constants $k$ and $k'$. 
These were fixed such that the quantized form of the two oscillators Hamiltonian does not comprise cross terms like $a^{\dagger 2}+a^2$ and $b^{\dagger 2}+b^2$.
In the new representation the quantized Hamilton operator looks like:
\begin{eqnarray}
\hat{H}&=&{\cal H}_{min}+\omega_1(a^{\dagger}a+\frac{1}{2})+\omega_2(b^{\dagger}b+\frac{1}{2})\nonumber\\
       &+&A_3kk'(a^{\dagger}b^{\dagger}+ba-a^{\dagger}b-b^{\dagger}a)-IjA_2\frac{1}{kk'}(a^{\dagger}b^{\dagger}+ba+a^{\dagger}b+b^{\dagger}a).
\end{eqnarray}
The off-diagonal terms will be treated by the equation of motion method. Thus, we have:
\begin{eqnarray}
&&[\hat{H},a^{\dagger}]=\omega_1a^{\dagger}+A_3kk'(b-b^{\dagger})-IjA_2\frac{1}{kk'}(b+b^{\dagger}),\nonumber\\
&&[\hat{H},b^{\dagger}]=\omega_2b^{\dagger}+A_3kk'(a-a^{\dagger})-IjA_2\frac{1}{kk'}(a+a^{\dagger}),\nonumber\\
&&[\hat{H},a]=-\omega_1a-A_3kk'(b^{\dagger}-b)+IjA_2\frac{1}{kk'}(b^{\dagger}+b),\nonumber\\
&&[\hat{H},b]=-\omega_2b-A_3kk'(a^{\dagger}-a)+IjA_2\frac{1}{kk'}(a^{\dagger}+a),
\end{eqnarray}
This is a linear system of equations which can be analytically solved. Indeed, one defines the phonon operator
\begin{equation}
\Gamma^{\dagger}=X_1a^{\dagger}+X_2b^{\dagger}-Y_1a-Y_2b,
\end{equation}
with the amplitudes $X_1, X_2, Y_1, Y_2$ fixed such that the following restrictions are satisfied:
\begin{equation}
\left[H,\Gamma^{\dagger}\right]=\Omega \Gamma^{\dagger},\;\;\left[\Gamma,\Gamma^{\dagger}\right]=1.
\end{equation}
The first restriction provides a homogeneous linear system of equations for the unknown amplitudes. The compatibility condition for this system leads to the equation (2.19) defining the phonon energy
$\Omega$.

\section{Appendix B}
\renewcommand{\theequation}{B.\arabic{equation}}
\setcounter{equation}{0}
Here we study the stability of the wobbling frequencies as given by Eq. (2.19). For what follows it is useful to introduce the notations:
\begin{eqnarray}
D_1&=&(2I-1)(\bar{A}_3-\bar{A}_1)+2j\bar{A}_1;\;\;F_1=(2j-1)(\bar{A}_3-\bar{A}_1)+2I\bar{A}_1,\nonumber\\
D_2&=&(2I-1)(\bar{A}_2-\bar{A}_1)+2j\bar{A}_1;\;\;F_2=(2j-1)(\bar{A}_2-\bar{A}_1)+2I\bar{A}_1\nonumber\\
G_1&=&\frac{2j-1}{j(j+1)}\sqrt{3}\left(\sqrt{3}\cos\gamma+\sin\gamma\right);\;\;G_2=\frac{2j-1}{j(j+1)}2\sqrt{3}\sin\gamma .
\end{eqnarray}
where $\bar{A}_k=A_k{\cal I}_0$.
We shall study the necessary conditions to obtain vanishing solutions for the equation:

\begin{equation}
\Omega^4+B\Omega^2+C=0,
\label{ecOm}
\end{equation}
{\bf The case}\;\;\;$C=0,\;\; B>0$.

From the equation defining $C$ one derives:
\begin{eqnarray}
&&D_1F_1+D_1G_1S-4Ij\bar{A}_3^2=0,\nonumber\\
&&D_2F_2+D_2G_2S-4Ij\bar{A}_2^2=0,
\end{eqnarray}
where  $S={\cal I}_0V$.

These are satisfied by:
\begin{equation}
S_1=\frac{4Ij\bar{A}_3^2-D_1F_1}{D_1G_1},\;\;\;S_2=\frac{4Ij\bar{A}_2^2-D_2F_2}{D_2G_2}
\end{equation}
{\bf The case} $B=0,\;\; C=\rm{arbitrary}$

The equation B=0 may be written in a different form:
\begin{equation}
G_1G_2S^2+(G_1+G_2)S+D_1D_2+8\bar{A}_1\bar{A}_2Ij+F_1F_2=0.
\end{equation}
We have the solutions:
\begin{eqnarray}
S_3&=&\frac{1}{2G_1G_2}\left(-(G_1+G_2)+\left[(G_1+G_2)^2-4G_1G_2\left(D_1D_2+8\bar{A}_1\bar{A}_2Ij+F_1F_2\right)\right]^{1/2}\right),\nonumber\\
S_4&=&\frac{1}{2G_1G_2}\left(-(G_1+G_2)-\left[(G_1+G_2)^2-4G_1G_2\left(D_1D_2+8\bar{A}_1\bar{A}_2Ij+F_1F_2\right)\right]^{1/2}\right).
\end{eqnarray}
Equations for $S_1, S_2, S_3, S_4$ represent four hyperbolic branches, in the plot of  ${\cal I}_0$ vs $V$.
\vskip1cm
These four curves, called separatrices,  define distinct regions in the parameters space, named nuclear phases,  where the system have specific properties. The equations of motion describe closed periodic curves surrounding the minimum point. When we approach a separatrice, the period tends to infinity which reflects the fact that the phase transition cannot take place.

\section{Appendix C}
\renewcommand{\theequation}{C.\arabic{equation}}
\setcounter{equation}{0}

The  operators used for the E2 and M1 transitions are:
\begin{eqnarray}
{\cal M}(E2,\mu)&=&e_{eff}\left[Q_0D^2_{\mu0}-Q_2(D^2_{\mu2}+D^2_{\mu-2})\right]
+e_{eff}\sum_{\nu}D^2_{\mu\nu}Y_{2\nu}r^2,\nonumber\\
 {\cal M}(M1,\mu)&=&\sqrt{\frac{3}{4\pi}}\mu_N\sum_{\nu=0,\pm1}\left[g_RI_{\nu}+(g_l-g_R)j_{\nu}+(g_s-g_l)s_{\nu}\right]D^1_{\mu\nu}\equiv M^{coll}_{1\mu}+M^{sp}_{1\mu}.\nonumber\\
\end{eqnarray}
where
\begin{equation}
Q_0=\frac{3}{4\pi}ZR_0^2\beta\cos\gamma,\;\;
Q_2=\frac{3}{4\pi}ZR_0^2\beta\sin\gamma/\sqrt{2}.
\end{equation}
The relative sign of the two terms involved in the expression of the E2 transition operator is compatible with the structure of the moments of inertia ${\cal J}^{rig}$ as well as with that of the single particle potential. Z and $R_0$ denote the nuclear charge and radius respectively, while $e_{eff}$ is the effective charge. Standard notations are used for  the nuclear magneton ($\mu_N$),the gyromagnetic factors of the rigid rotor ($G_R=Z/A$) and single particle, characterizing the  orbital angular momentum 
($g_l$) and the spin ($g_s$), respectively. 

The reduced E2 and M1 transition probabilities have the expressions:
\begin{eqnarray}
B(E2;I_i^{\pi};jn_w\to I_f^{\pi};jn^{\prime}_w)&=&\langle I_i^{\pi};jn_w||{\cal M}(E2)||I_f^{\pi};jn^{\prime}_w\rangle ^2,\nonumber\\
B(M1;I_i^{\pi};jn_w\to I_f^{\pi};jn^{\prime}_w)&=&\langle I_i^{\pi};jn_w||{\cal M}(M1)||I_f^{\pi};jn^{\prime}_w\rangle ^2.
\end{eqnarray}
Note that the reduced matrix elements are defined according to the Rose convention \cite{Rose}:
\begin{equation}
\langle JM|T_{k\mu}|J'M'\rangle=C^{J'\;k\;J}_{M'\;\mu\;M}\langle J||T_{k}||J'\rangle.
\end{equation}
The reduced matrix elements for the electric and magnetic transition operators have the analytical expressions given in Appendix D of Ref. \cite{Rad017}.  The mixing ratio for the E2 and M1 transitions from the TSD2 to TSD1 bands, is defined as \cite{Toki,Kran}:
\begin{equation}
\delta=8.78\times 10^{-4}E_{if}\frac{\langle I_i||{\cal M}(E2)||I_f\rangle}{\langle I_i||{\cal M}(M1)||I_f\rangle}.
\end{equation}
The matrix element for the E2 transition is taken in units of ${\rm e\cdot fm^2}$, while that for the M1 transition in ${\rm e\cdot fm}$. The transition energy is denoted by $E_{if}$ and is taken in MeV.
We also calculated the transition quadrupole moments according to the definitions \cite{Hage1}:
\begin{equation}
Q_I=\sqrt{\frac{16\pi}{5}}\langle I||{\cal M}(E2)||I-2\rangle / C^{I\;2\;I-2}_{K\;0 K}.
\end{equation}
The transition quadrupole moment depends only weakly on the K quantum number. In our calculations one used $K=\frac{1}{2}$.
The calculated matrix elements involve an effective charge $e_{eff}=1.5$.  Results of our numerical analysis are described in Section IV.

\end{document}